# Direct observation of the magnetic ground state of the two smallest triangular nanographenes

*Elia Turco,*[⊥] *Annika Bernhardt,*[||] *Nils Krane,*[⊥] *Leoš Valenta,*[||] *Roman Fasel,*[⊥,§] *Michal Juríček,*[||,*] *and Pascal Ruffieux,*[⊥,*]

⊥ nanotech@surfaces Laboratory, Empa − Swiss Federal Laboratories for Materials Science and Technology, 8600 Dübendorf, Switzerland

|| Department of Chemistry, University of Zurich, Winterthurerstrasse 190, 8057 Zurich, Switzerland

§ Department of Chemistry, Biochemistry and Pharmaceutical Sciences, University of Bern, 3012 Bern, Switzerland

ABSTRACT

Fusion of three benzene rings in a triangular fashion gives rise to the smallest open-shell graphene fragment, the phenalenyl radical, whose π-extension leads to an entire family of non-Kekulé triangular nanographenes with high-spin ground states. Here, we report the first synthesis of unsubstituted phenalenyl on a Au(111) surface, which is achieved by combining in-solution synthesis of the hydro-precursor and on-surface activation by atomic manipulation, using the tip of a scanning tunneling microscope (STM). Single-molecule structural and electronic characterization confirm its open-shell $S = ½$ ground state that gives rise to Kondo screening on the Au(111) surface. In addition, we compare the phenalenyl's electronic properties with those of triangulene, the second homolog in the series, whose $S = 1$ ground state induces an underscreened Kondo effect. Our results set a new lower size-limit in the on-surface synthesis of magnetic nanographenes that can serve as building blocks for the realization of new exotic quantum phases of matter.

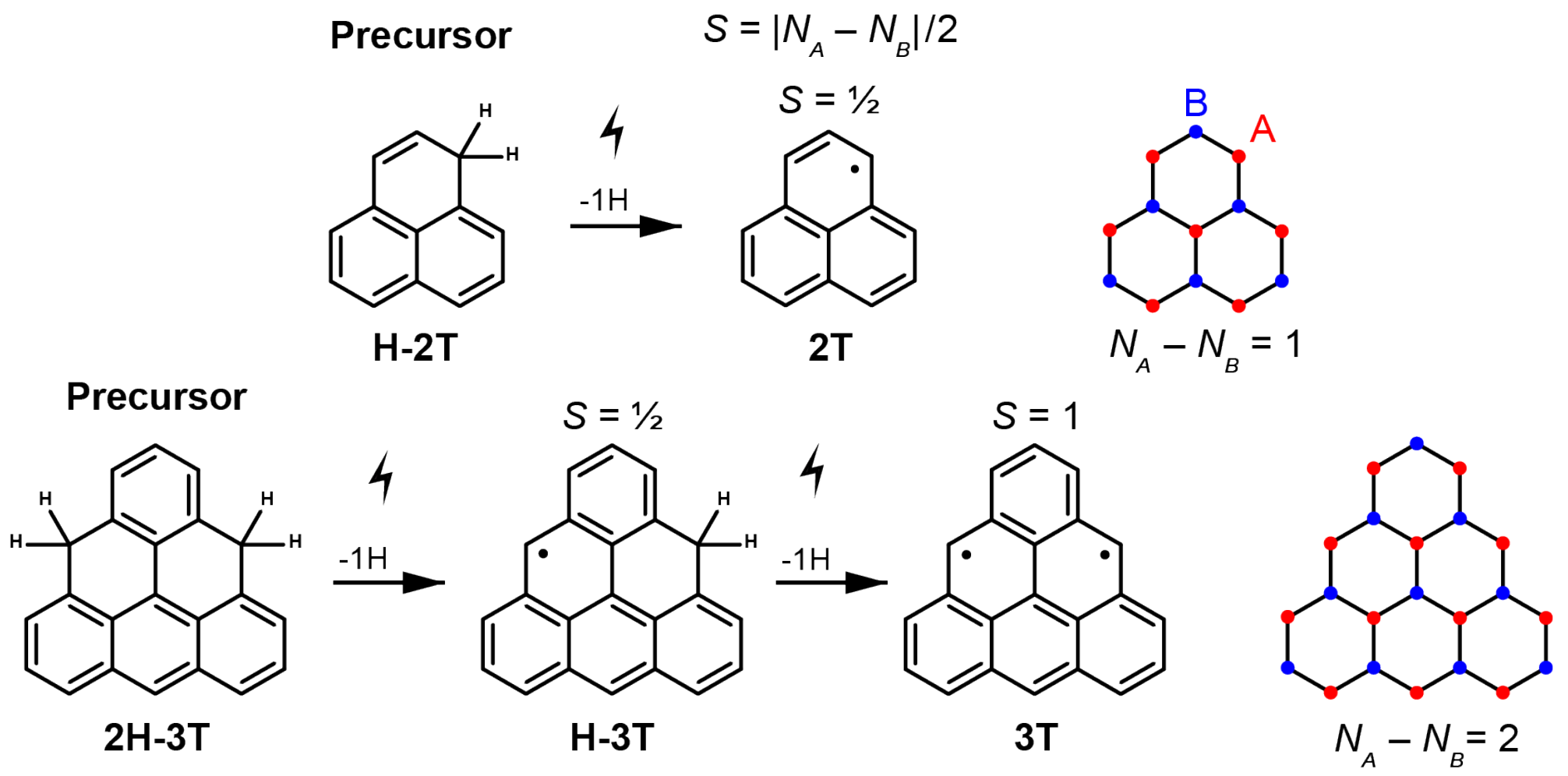

**Figure 1.** Synthesis of 2T and 3T from the corresponding hydro-precursors. Removal of one hydrogen atom from each of the $sp^3$-hybridized carbon atoms leads to non-Kekulé polycyclic conjugated hydrocarbons 2T and 3T, where the imbalance between sublattices $N_A$ and $N_B$ determines the total spin quantum number $S$ according to Ovchinnikov's rule[1].

The bipartite nature of graphene's honeycomb lattice offers the opportunity to design open-shell nanographenes (NGs) with tailor-made magnetic ground states. For specific NG topologies (also known as non-Kekulé[2]), the imbalance between the two interpenetrating lattices A and B leads to the presence of unpaired electrons forming non-trivial magnetic ground states with a total spin $S = |N_A - N_B|/2$, as predicted by Ovchinnikov and Lieb about 50 years ago[1,3]. Unlike localized *d*- or *f*-shell electrons, π-conjugated radicals are highly delocalized and prone to interact with neighboring unpaired electrons to form strongly correlated magnetic states, which are the main requisite for measurement-based quantum computation[4]. In this regard, the family of zigzag-edged triangular NGs represent a prototypical class of polybenzenoid magnetic building blocks for the realization of such entangled spin systems. These odd-alternant triangular graphene fragments, commonly denoted as [*n*]triangulenes or simply *n*T, where $n \geq 2$ is the number of benzene rings per edge, possess a total spin $S$ that scales with triangulene size[5]. Since the early works of Clar[6], the two smallest mono- and diradical polycyclic conjugated hydrocarbons 2T and 3T have played a central role in the fundamental understanding of the reactivity and electronic properties of open-shell compounds. In 1957, Calvin[7] described for the first time the phenalenyl radical (2T), formed by coincidence from phenalene by oxidation, where the presence of the radical species was confirmed by EPR spectroscopy. However, the high spin density symmetrically distributed over the majority sublattice periphery (α-positions) makes zigzag edges of 2T highly reactive and subject to σ-dimerization and oxidation in air[8], preventing the isolation and characterization of the pristine compound. Steric protection of the reactive sites and thermodynamic stabilization *via* π-extension allowed the solution-based synthesis and characterization of persistent 2T[9–12] and more recently 3T derivatives[13–15]. On-surface synthesis under ultrahigh vacuum conditions turned out to be an efficient route towards the realization of unsubstituted open-shell compounds[16,17] whose structural and electronic properties can be characterized *in situ* at the single-molecule level by means of scanning probe techniques. With this approach, the synthesis of unsubstituted 3T was successfully achieved on Cu(111), NaCl(100) and Xe(111) using the tip of a scanning probe microscope[18]. Since then, the triangulene family has grown rapidly with the on-surface synthesis of even more challenging multiradical 4T[19], 5T[20] and 7T[21]. Just recently, these magnetic building blocks were coupled into dimers[22], trimers[23], rings[24] and 1D chains[25] revealing first examples of fascinating correlated magnetic ground states.

Despite the rich in-solution and on-surface synthesis of triangular NGs, the pristine structure and electronic properties of the smallest member of the family, namely, 2T, have not yet been explored. Here, we report the synthesis of 2T and 3T following a combined in-solution and on-surface activation approach. The chosen strategy involves in-solution synthesis of non-reactive hydro-precursors, H-2T and 2H-3T, that are deposited on a metal surface under ultrahigh vacuum conditions and subsequently activated to 2T and 3T by tip-induced dehydrogenation. Their *in situ* characterization by scanning tunneling microscopy (STM) and spectroscopy (STS) measurements yields a direct proof of the magnetic ground state of 2T and 3T on Au(111) *via* the observation of a Kondo resonance.

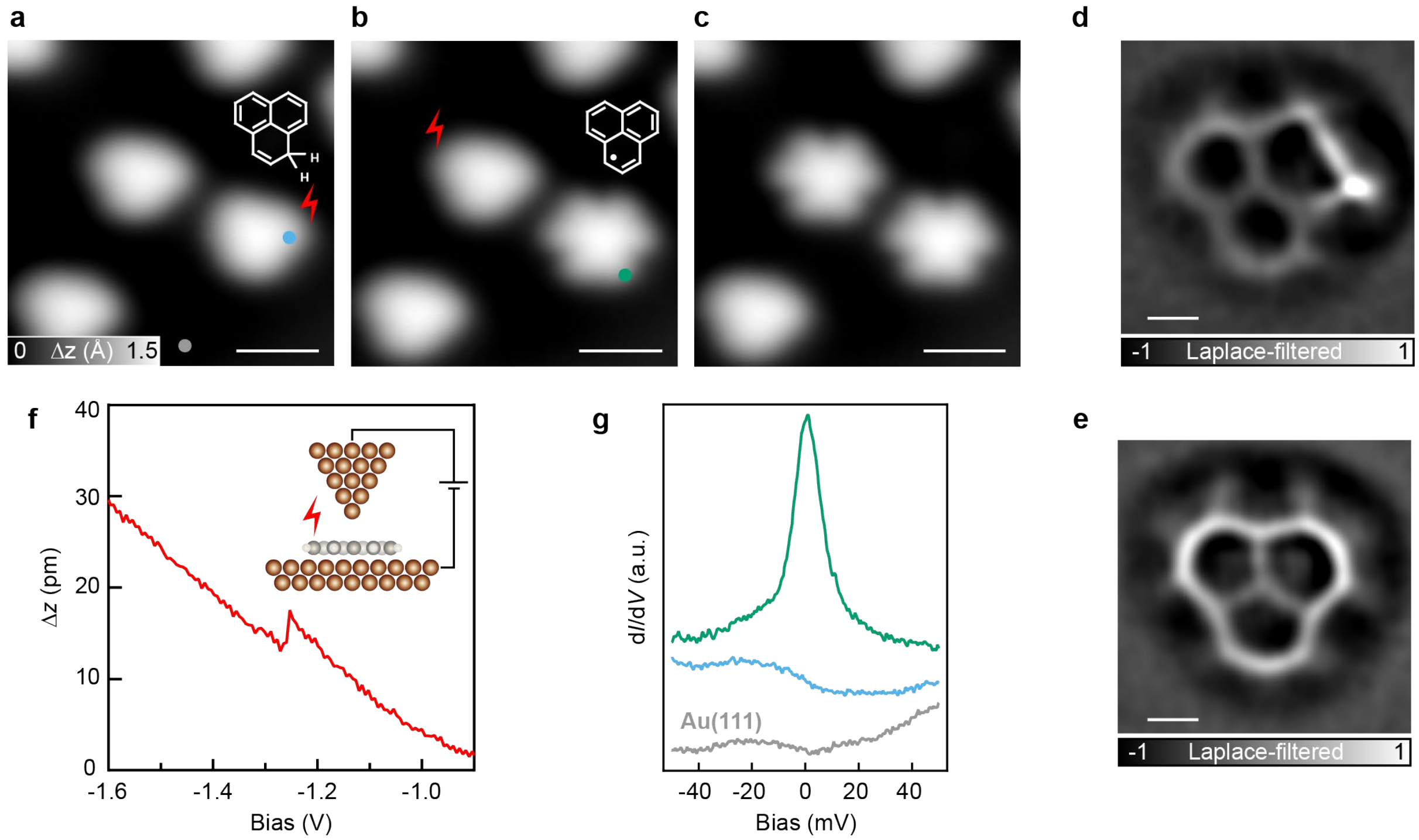

**Figure 2.** Tip-induced activation of H-2T on Au(111). (**a**,**b**,**c**) Series of STM images showing the selective dehydrogenation of H-2T to 2T ($V$ = –0.6 V, $I$ = 100 pA). (**d**,**e**) Laplace-filtered nc-AFM images of H-2T and 2T molecules. Open feedback parameters: $V$ = –50 mV, $I$ = 100 pA; $\Delta z$ = –1.8 Å (H-2T) and –1.5 Å (2T). (**f**) Cleaving of the hydrogen atom is detected as a step in constant-current $z(V)$ spectroscopy ($I$ = 50 pA). (**g**) Constant-height d$I$/d$V$ spectra acquired on H-2T and 2T, revealing the Kondo screening of the radical in 2T. Scale bars: 1 nm (a,b,c), 0.2 nm (d,e).

## RESULTS AND DISCUSSION

**In-solution synthesis of H-2T and 2H-3T**

H-2T was prepared from commercially available 3-(naphthalen-1-yl)propanoic acid in three steps. The phenalene core was built up in a Friedel–Crafts acylation. The reduction of 2,3-dihydro-1*H*-phenalen-1-one and a subsequent dehydration reaction yielded H-2T. The air-sensitive compound was purified by column chromatography and subsequent sublimation. The experimental details of the in-solution synthesis of H-2T are reported in the Supplementary Information.

The synthesis of 2H-3T was achieved by combining the procedures of Šolomek and Juríček[26] with that of Johnson[27]. This approach makes this precursor accessible in only two steps with an overall yield of 70%. The route employs commercially available *o*-bromobenzyl alcohol that is reacted with two equivalents of *n*-butyllithium before it is quenched with diethyl carbonate to form a tetraalcohol intermediate. The cyclization of the tetraalcohol with triflic acid provides triangulenyl cation, an unstable intermediate that is reduced in situ, resulting in a mixture of two isomers of 2H-3T, as described previously[26,27].

**On-surface synthesis of 2T and 3T**

The hydro-precursors H-2T and 2H-3T were sublimed under ultrahigh vacuum conditions onto a Au(111) surface held at room temperature. STM imaging of the resulting surface (reported in Fig. S1) reveals that both precursors are adsorbed as individual molecules that are predominantly located in the face-centered cubic (fcc) regions of the herringbone reconstruction of Au(111) when deposited at submonolayer coverage. Together with the target hydro-precursors, we also observed a minority of different species, often self-assembled into molecular clusters, that we assign to oxidized molecules[18]. Despite the presence of the additional hydrogen atom, which breaks the three-fold symmetry, STM imaging of H-2T species (Fig. 2a) reveals an almost uniform triangular apparent shape, while its nonplanar chemical structure is directly resolved by bond-resolved non-contact atomic force microscopy (nc-AFM)[28] (Fig. 2d).

Dehydrogenation of H-2T and 2H-3T, i.e., the removal of one hydrogen atom from the $sp^3$-hybridized carbon atoms, was achieved by means of atomic manipulation[29,30]. Specifically, we positioned the STM/nc-AFM tip above one of the corners of H-2T and, while keeping the current fixed to few tens of pA, increased the voltage until a jump in tip height was observed (Fig. 2f). Subsequent STM imaging (Fig. 2b) revealed a significant change in the apparent shape of the target molecule, while all the other molecules remained unaffected. The successful dehydrogenation of H-2T was confirmed by nc-AFM images such as shown in Fig. 2e, which reveal a stable and flat geometry of 2T on the Au(111) surface. In addition to STM/nc-AFM imaging, we also followed the dehydrogenation of H-2T by low-energy differential conductance ($dI/dV$) spectroscopy. Figure 2g shows a clear transition from a featureless low-energy conductance spectrum for the H-2T precursor to a sharp zero-bias resonance measured at the six equivalent lobes located at the edges of 2T. This zero-bias resonance is the characteristic fingerprint of the Kondo screening of a spin $S = ½$ impurity on a metal substrate, thus attesting to the open-shell ground state of the 2T species, as discussed in more detail below. We repeated the tip-based activation procedure of the precursor H-2T into 2T with different metal tips and over 50 different molecules, which revealed that dehydrogenation is bias voltage-dependent and occurs on average at a sample bias of –1.3 V. An alternative way of dehydrogenating H-2T on Au(111) consists of annealing the sample to 180°C (see Fig. S2 for details). In an analogous way to activating H-2T into 2T, triangulene 3T was generated from the dihydro-precursor 2H-3T by a tip-induced dehydrogenation of the two $sp^3$-hybridized carbon atoms, as will be discussed later.

## Electronic structure of 2T

The bond topology of 2T intrinsically leads to the presence of an unpaired electron, which in the tight-binding (TB) model is depicted as a non-bonding, half-filled zero-energy state. If we also consider the on-site electron–electron Coulomb repulsion $U$ within the mean-field Hubbard (MFH) level of theory, this zero-energy state splits into a singly occupied and a singly unoccupied molecular orbital (SOMO/SUMO) with opposite spin orientations (Fig. 3a).

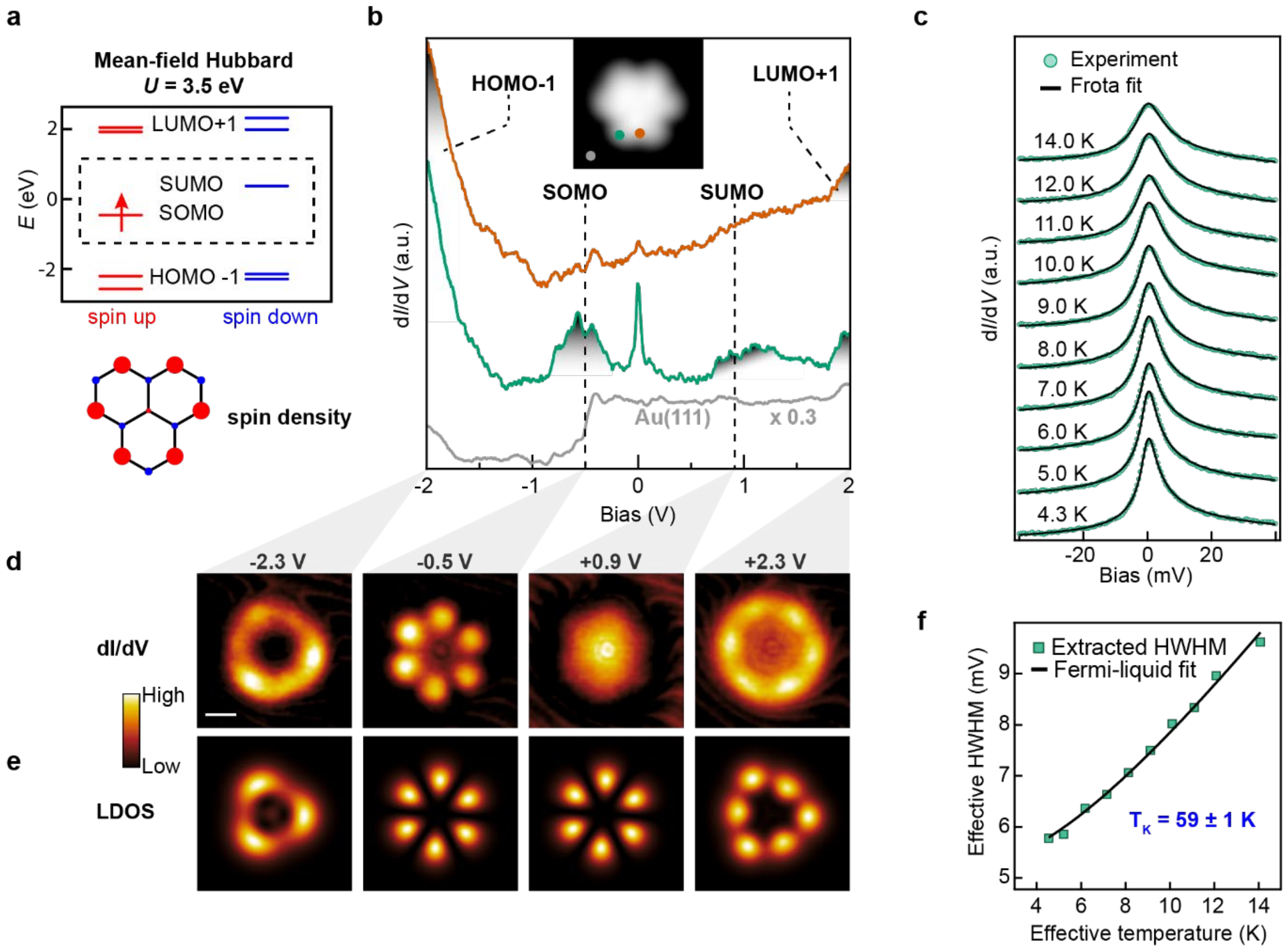

**Figure 3.** Electronic characterization of 2T. (**a**) MFH energy diagram of 2T, where $U$ denotes the on-site Coulomb repulsion. Along with the energy spectrum, the spin density distribution is depicted, where red- and blue-filled circles denote mean populations of spin-up and spin-down electrons, respectively. (**b**) d$I$/d$V$ spectroscopy on 2T acquired with a CO-functionalized tip reveals four molecular orbital resonances (open feedback parameters: $V$ = –2.0 V, $I$ = 250 pA; $V_{rms}$ = 16 mV). Acquisition positions are indicated in the STM image shown in the inset. (**c**) Kondo resonance of 2T as a function of sample temperature (open feedback parameters: $V$ = –50 mV, $I$ = 1 nA; $V_{rms}$ = 0.4 mV). (**d**,**e**) Constant-current d$I$/d$V$ maps of the HOMO-1, SOMO, SUMO and LUMO+1 resonances of 2T (d), along with the corresponding MFH-LDOS maps (e). Tunneling parameters for the d$I$/d$V$ maps: $I$ = 300 pA, $V_{rms}$ = 24 mV. Scale bar: 0.5 nm. (**f**) Half-width a half-maximum (HWHM) of the d$I$/d$V$ spectra in (c) extracted from the Frota fits and plotted versus the effective temperature, as described in Ref. [16]. The solid black line shows the best fit with the function $\frac{1}{2}\sqrt{\left(\alpha k_B T_{eff}\right)^2 + (2k_B T_K)^2}$, which describes the thermal broadening of the Kondo resonance in the Fermi liquid model[31]. Fitting parameters: $T_K \sim 59$ K and $\alpha = 13$.

The unpaired electron giving rise to the net spin $S = ½$ of 2T is delocalized over the entire molecular framework, but with the highest probability density at the six α-positions, as shown by the spin density plot in Fig. 3a. The TB-MFH theoretical predictions are entirely confirmed by our scanning tunneling spectroscopy (STS) measurements of isolated 2T molecules on Au(111). The differential conductance d$I$/d$V$ spectra, acquired at two different positions on the molecule, reveal four distinct peaks in the local density of states (LDOS), at –2.0 V, –0.5 V, +0.9 V and 2.0 V (Fig. 3b). To assign these resonances to the calculated MFH-LDOS, shown in Fig. 3e, we did a spatial mapping of each d$I$/d$V$ resonance (Fig. 3d). The excellent match between the calculated LDOS and experimental d$I$/d$V$ maps proves the correct assignment of the molecular orbitals (MOs). As indicated in Fig. 3d and 3e, SOMO and SUMO share the same LDOS. Indeed, for a molecule with a singly occupied orbital, tunneling at opposite bias polarities involves adding/removing an electron to/from this orbital[33]. The Coulomb energy penalty for doubly occupying this orbital with respect to the unoccupied case, gives rise to the experimentally observed energy gap of 1.4 eV. Despite the identical symmetry of the experimental SOMO and SUMO d$I$/d$V$ maps, the energetic overlap of the SUMO state with the gold surface state hinders its clear visualization. This can be improved by performing d$I$/d$V$ mapping with a CO-terminated tip, which reveals an increased sensitivity to the SUMO (Fig. S3). However, a much more direct proof for the presence of an unpaired spin on a metal surface is the detection of a zero-bias resonance (ZBR) in the d$I$/d$V$ spectrum that is related to the Kondo screening of its magnetic moment[34,35]. Spatial mapping of this ZBR (Fig. 4e) indirectly gives access to the spin density distribution, which naturally resembles the LDOS of the singly occupied/unoccupied MOs. To confirm that the ZBR peak indeed derives from a Kondo resonance, we determined the ZBR linewidth as a function of sample temperature (Fig. 3c). It is found to broaden non-thermally, and to follow the characteristic trend of a $S = ½$ Kondo-screened state with a Kondo temperature $T_K = 59 \pm 1$ K (Fig. 3f).

**Kondo screening in *S* = ½ and *S* =1 triangular NGs**

The second point that we want to address is a direct comparison of the Kondo screening in the $S = ½$ and $S = 1$ NGs 2T and 3T, respectively. The first evidence of the ferromagnetic $S = 1$ ground state of 3T was reported by Pavliček et al. on non-metallic surfaces[18]. Here, we provide direct evidence that 3T retains its triplet magnetic ground state when adsorbed on Au(111). A detailed electronic characterization of 3T is reported in Fig. S5, while here we will focus on the low-energy magnetic properties. As already shown for 2T, the presence of a magnetic impurity with $(2S + 1)$-degenerate spin ground state coupled to the electron bath of the substrate results in a (complete or partial) screening of the magnetic moment of the impurity. If we have a single screening channel (the conduction band of the Au substrate), the spin of the magnetic adsorbate is screened from a spin $S$ to an effective spin $(S - ½)$[36,37]. This implies that, for systems with $S > ½$, the spin is not fully screened, and one thus speaks of an *underscreened* Kondo effect, where the residual magnetic moment has a Zeeman energy much smaller than the Kondo temperature[38,39]. Here, by co-depositing both H-2T and 2H-3T onto the Au(111) surface, we investigate the differences between a fully screened ($S = ½$, 2T) and an underscreened ($S = 1$, 3T) Kondo state. The high-resolution STM image in Fig. 4a shows the typical apparent shape of 2H-3T, where the two additional hydrogen atoms (one at each $sp^3$ center) "quench" the diradical nature, resulting in a featureless low-energy d$I$/d$V$ spectrum (Fig. 4d). Employing the aforementioned manipulation technique, we first selectively removed one hydrogen atom, inducing a clear modification in the apparent shape of the molecule (Fig. 4b). The so obtained H-3T molecule reveals a strong ZBR characteristic of a Kondo-screened $S = ½$ ground state. We then proceeded and removed the second hydrogen atom, to obtain the target compound 3T (Fig. 4c, left) featuring a three-fold symmetric apparent shape. In order to compare the Kondo-related ZBR of 2T and 3T with the same metal tip, we also activated a nearby 2T molecule (Fig. 4c, right) and performed low-bias STS on both 2T and 3T.

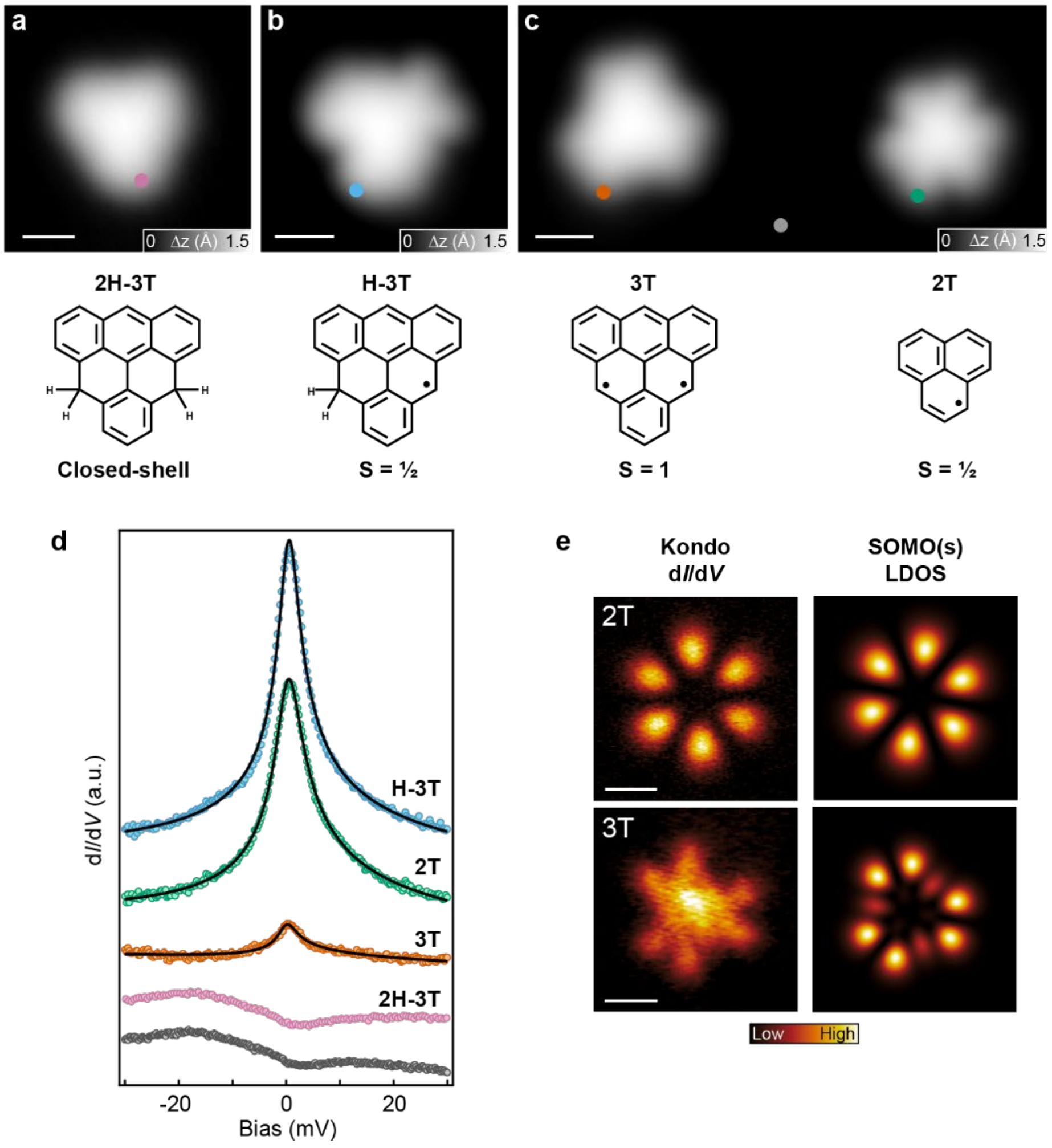

**Figure 4.** Comparison of spin-½ (2T) and spin-1 (3T) Kondo resonances. (**a**,**b**,**c**) High-resolution STM images of 2H-3T ($V = –0.1$ V, $I = 20$ pA), H-3T ($V = –0.1$ V, $I = 100$ pA), 3T and 2T ($V = –0.1$ V, $I = 100$ pA) obtained by successive tip-induced dehydrogenation. (**d**) Low-bias d$I$/d$V$ spectra (colored dots) acquired on (a,b,c) using a metal tip (open feedback parameters: $V =$ –50 mV, $I = 1$ nA; $V_{rms} = 0.35$ mV), revealing a ZBR for the $S = ½$ and $S = 1$ systems. The positions of the acquired spectra are indicated in (a,b,c). The d$I$/d$V$ background from the Au(111) substrate surface (grey curve) was subtracted from the 3T, 2T and H-3T d$I$/dV spectra. Frota fits are shown as black lines (see Fig. S7 for details). (**e**) Experimental constant-height d$I$/d$V$ maps of the ZBR in 2T and 3T (left panels), with the corresponding MFH-TB LDOS SOMO(s) of 2T and 3T (right panels). A subtraction of the off-resonance d$I$/d$V$ maps ($V$ = 30 mV) was applied to remove the uniform d$I$/d$V$ background of the molecules. Open feedback parameters; 2T: $V$ = –100 mV, $I$ = 1 nA; $V_{rms}$ = 1 mV, 3T: $V$ = –100 mV, $I$ = 1.4 nA; $V_{rms}$ = 2 mV). Scale bars: 0.5 nm (a,b,c,e).

The low-energy differential conductance spectra shown in Fig. 4d reveal a pronounced intensity difference of the ZBRs of 2T and 3T. From a Frota fit[40,41] to the 2T and 3T spectra, we determined an amplitude ratio of 7 to 1. This remarkable difference is consistent with recent observations on similar $S$ = ½ and $S$ = 1 systems[42–44], where significantly reduced intensity for underscreened Kondo resonances was reported. According to early experiments on $d$-electron materials[45] and more recent renormalization group studies within the Anderson model[46], the Hund coupling of the spins in a $S$ > ½ ground state determines a quenching of the effective Kondo coupling, which results in an exponentially reduced Kondo temperature compared to a S = ½ Kondo system. Considering the strong Hund coupling of hundreds of meV in 3T[47,48] and the recent experimental results on similar $S$ = 1 nanographenes, we expect a Kondo temperature within our limited experimentally accessible temperature range (4.5 K < $T$ < 14 K). Therefore, a correct estimation of the Kondo temperature would require a wider range in temperatures and a tunable magnetic field[49,50], which is beyond our current experimental capabilities. Nevertheless, since the ZBRs of 2T and 3T were measured in the very same conditions, we expect that a lower Kondo temperature would correspond to a narrower half width at half maximum (HWHM). Frota fitting of 2T and 3T spectra revealed that the ZBR of 3T is significantly narrower than 2T ZBR, supporting the aforementioned hypothesis of a lower Kondo temperature of 3T. More details on the HWHMs of each ZBR are reported in Fig. S7.

In summary, we have reported the synthesis of unsubstituted [2]triangulene (phenalenyl radical, 2T) and [3]triangulene (3T) on Au(111) via tip-induced dehydrogenation of hydro-/dihydro-precursors, respectively. Single-molecule STM/nc-AFM and STS measurements have provided a comprehensive characterization of their chemical structure as well as their electronic and magnetic properties. A direct proof for their open-shell $S$ = ½ and $S$ = 1 ground state, respectively, has been obtained by the observation of Kondo screening of the unpaired spin(s) by the Au substrate electrons. Notably, the ferromagnetically coupled spins in triangulene 3T give rise to an underscreened Kondo effect, as recently predicted by multiorbital Anderson impurity model calculations[51]. On the other hand, the phenalenyl radical 2T with its three-fold symmetry and spin-½ doublet ground state constitutes a prototypical all-carbon magnetic building block, whose successful on-surface synthesis and characterization opens new opportunities for the bottom-up synthesis of strongly correlated one- and two-dimensional carbon-based spin-chains and lattices.

ASSOCIATED CONTENT

**Supporting Information.**

Experimental and computational methods, supporting STM and STS data, additional calculations, Frota fitting parameters, and a detailed synthetic description of chemical compounds reported in this study.

The Supporting Information is available free of charge at http://pubs.acs.org.
The following files are available.
Supplementary Information (PDF)

**Corresponding Authors**

michal.juricek@chem.uzh.ch

pascal.ruffieux@empa.ch

**Author Contributions**

P.R., R.F. and M.J. conceived the experiments. A.B and L.V synthesized and characterized the precursors in solution. E.T and N.K performed the on-surface synthesis and scanning probe measurements. E.T performed the tight-binding calculations and analyzed the data. All authors discussed the results and contributed to the writing of the manuscript.

ACKNOWLEDGMENT

This work was supported by the Swiss National Science Foundation (grant no. 200020-182015 and CRSII5_205987, M.J./PP00P2_170534 and PP00P2_198900), the NCCR MARVEL funded by the Swiss National Science Foundation (grant no. 51NF40-182892), the EU Horizon 2020 research and innovation program – Marie Skłodowska-Curie grant no. 813036 and Graphene Flagship Core 3 (grant no. 881603), the Office of Naval Research (grant no. N00014-18-1-2708), ERC Consolidator grant (T2DCP, grant no. 819698), ERC Starting grant (M.J./INSPIRAL, grant no. 716139). This work was supported by a grant from the Swiss National Supercomputing Centre (CSCS) under project ID s1141. We thank Carlo A. Pignedoli, Kristjan Eimre and Ilias Gazizullin for fruitful scientific discussions. Skillful technical assistance by Lukas Rotach is gratefully acknowledged.

**Notes**

The authors declare no competing financial interest.

# Supporting Information

# Direct observation of the magnetic ground state of the two smallest triangular nanographenes

*Elia Turco,*[⊥] *Annika Bernhardt,*[||] *Nils Krane,*[⊥] *Leoš Valenta,*[||] *Roman Fasel,*[⊥,§] *Michal Juríček,*[||,*]

*and Pascal Ruffieux,*[⊥,*]

⊥ nanotech@surfaces Laboratory, Empa − Swiss Federal Laboratories for Materials Science and Technology, 8600 Dübendorf, Switzerland

|| Department of Chemistry, University of Zurich, Winterthurerstrasse 190, 8057 Zurich, Switzerland

§ Department of Chemistry, Biochemistry and Pharmaceutical Sciences, University of Bern, 3012 Bern, Switzerland

Corresponding Author:

michal.juricek@chem.uzh.ch

pascal.ruffieux@empa.ch

## Contents

## 1. Methods

### Sample Preparation and Scanning Probe Measurements

STM measurements were performed with a commercial low-temperature STM/AFM from Scienta Omicron operated at a temperature of 4.5 K and base pressure below $5\times10^{-11}$ mbar. The Au(111) single crystal surfaces were prepared by iterative $Ar^{+}$ sputtering and annealing cycles. Prior to sublimation of molecules, the surface quality was checked through STM imaging. The powders of H-2T and 2H-3T precursors were filled into quartz crucibles of a home-built evaporator and sublimed at 70 °C and 130 °C, respectively, onto the single crystal surfaces held at room temperature. STM images were acquired both in constant-current (overview and high-resolution imaging) and constant-height (bond-resolved imaging) modes, dI/dV spectra were acquired in constant-height mode and dI/dV maps were acquired in constant-current mode. Indicated bias voltages are given with respect to the sample. Unless otherwise noted, all measurements were performed with metallic tips. Differential conductance dI/dV spectra and maps were obtained with a lock-in amplifier. Modulation voltages (root mean square amplitude $V_{rms}$) for each measurement are provided in the respective figure caption. Bond-resolved nc-AFM (STM) images were acquired in constant-height mode with CO-functionalized tips at low bias voltages while recording frequency and current signals. Open feedback parameters on the molecular species and subsequent lowering of the tip height ($\Delta z$) for each image are provided in the respective figure captions. The data was processed with Wavemetrics Igor Pro software[1].

## Tight-binding and mean-field Hubbard calculations

TB-MFH calculations were performed by numerically solving the mean-field Hubbard Hamiltonian with third-nearest-neighbor hopping

$$\hat{H}_{MFH} = \sum_{j} \sum_{\langle\alpha,\beta\rangle_j,\sigma} -t_j c^{\dagger}_{\alpha,\sigma} c_{\beta,\sigma} + U \sum_{\alpha,\sigma} \langle n_{\alpha,\sigma}\rangle n_{\alpha,\bar{\sigma}} - U \sum_{\alpha} \langle n_{\alpha,\uparrow}\rangle\langle n_{\alpha,\downarrow}\rangle. \quad \text{(S1)}$$

Here, $c^{\dagger}_{\alpha,\sigma}$ and $c_{\beta,\sigma}$ denote the spin selective ($\sigma \in \{\uparrow,\downarrow\}$ with $\bar{\sigma} \in \{\downarrow,\uparrow\}$) creation and annihilation operator at sites $\alpha$ and $\beta$, $\langle\alpha,\beta\rangle_j$ ($j = \{1, 2, 3\}$) denotes the nearest-neighbor, second-nearest-neighbor and third-nearest-neighbor sites for $j$ = 1, 2 and 3, respectively, $t_j$ denotes the corresponding hopping parameters (with $t_1$ = 2.7 eV, $t_2$ = 0.1 eV and $t_3$ = 0.27 eV for nearest-neighbor, second-nearest-neighbor and third-nearest-neighbor hopping[3]), $U$ denotes the on-site Coulomb repulsion, $n_{\alpha,\sigma}$ denotes the number operator, and $\langle n_{\alpha,\sigma}\rangle$ denotes the mean occupation number at site $\alpha$. Orbital electron densities, $\rho$, of the $n^{\text{th}}$-eigenstate with energy $E_n$ have been simulated from the corresponding state vector $a_{n,i,\sigma}$ by

$$\rho_{n,\sigma}(\vec{r}) = \left| \sum_{i} a_{n,i,\sigma} \phi_{2p_z}(\vec{r} - \vec{r}_i) \right|^2, \quad \text{(S2)}$$

where $i$ denotes the atomic site index and $\phi_{2p_z}$ denotes the Slater $2p_z$ orbital for carbon.

All the TB-MFH calculations presented in the manuscript were done in the third-nearest-neighbor approximation and using an on-site Coulomb term $U$ = 3.5 eV.

The TB-MFH software library[4] was developed within the Python programming language. The code is open-source, available at https://github.com/eimrek/tb-mean-field-hubbard.

## 2. Supporting STM, STS and theoretical data

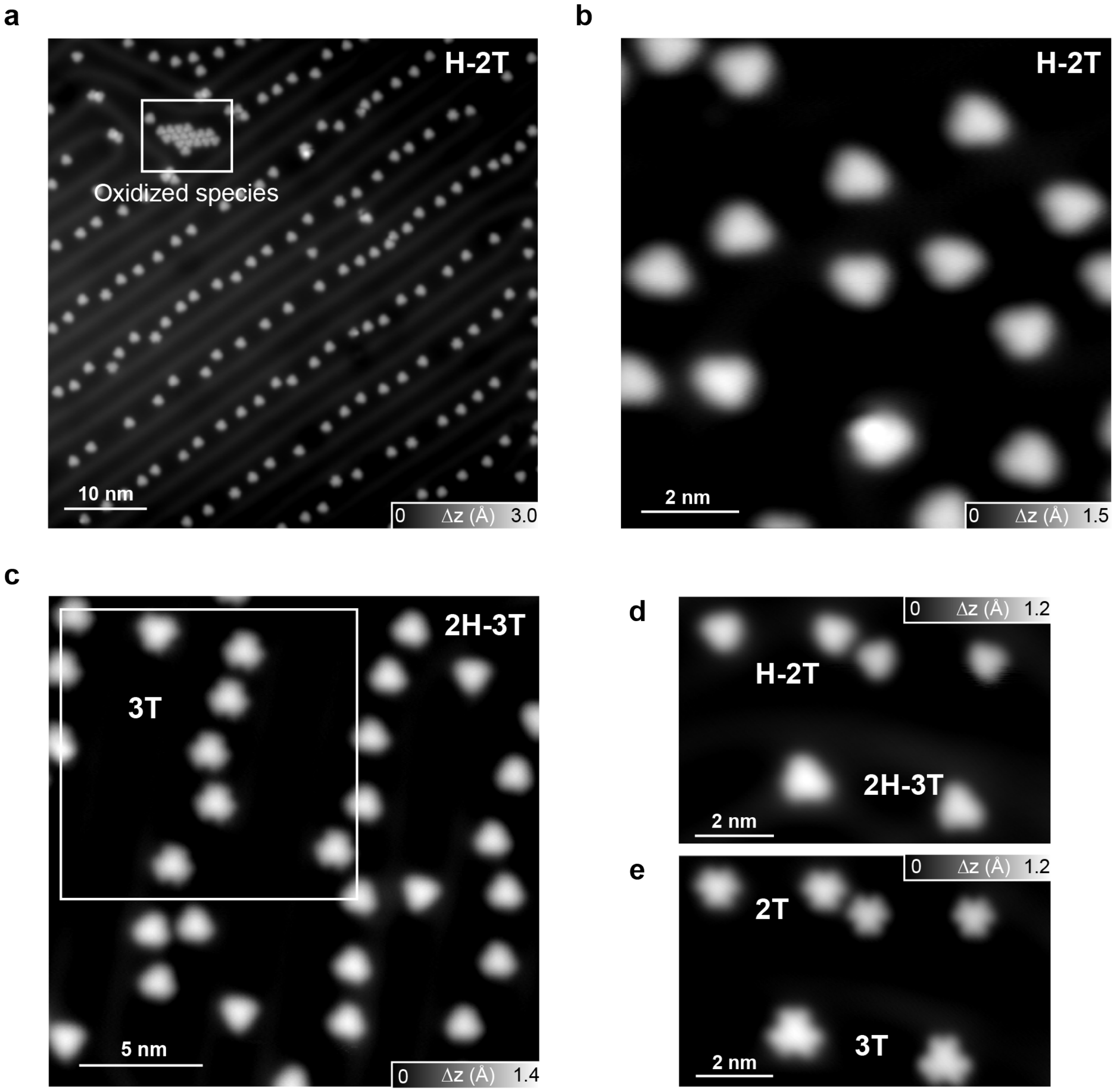

**Figure S1.** (**a**,**b**) Overview STM image of sub-monolayer coverage of H-2T as deposited on a Au(111) surface held at room temperature. The inset in (a) highlights the presence of self-assembled structures that we assign to oxidized species. (**c**) Overview STM image of sub-monolayer coverage of 2H-3T as deposited on a Au(111) surface at room temperature. The 3T molecules highlighted in the white frame were formerly activated by tip-induced dehydrogenation of 2H-3T precursors. (**d,e**) Sample with both H-2T and 2H-3T precursors, as deposited onto the Au(111) surface (d), and after the tip-based activation of all the molecular species (e). Scanning parameters: (a) $V = -1$ V, $I = 50$ pA; (b) $V = -0.8$ V, $I = 200$ pA; (c) $V = -0.6$ V, $I = 100$ pA; (d,e) $V = -1$ V, $I = 50$ pA.

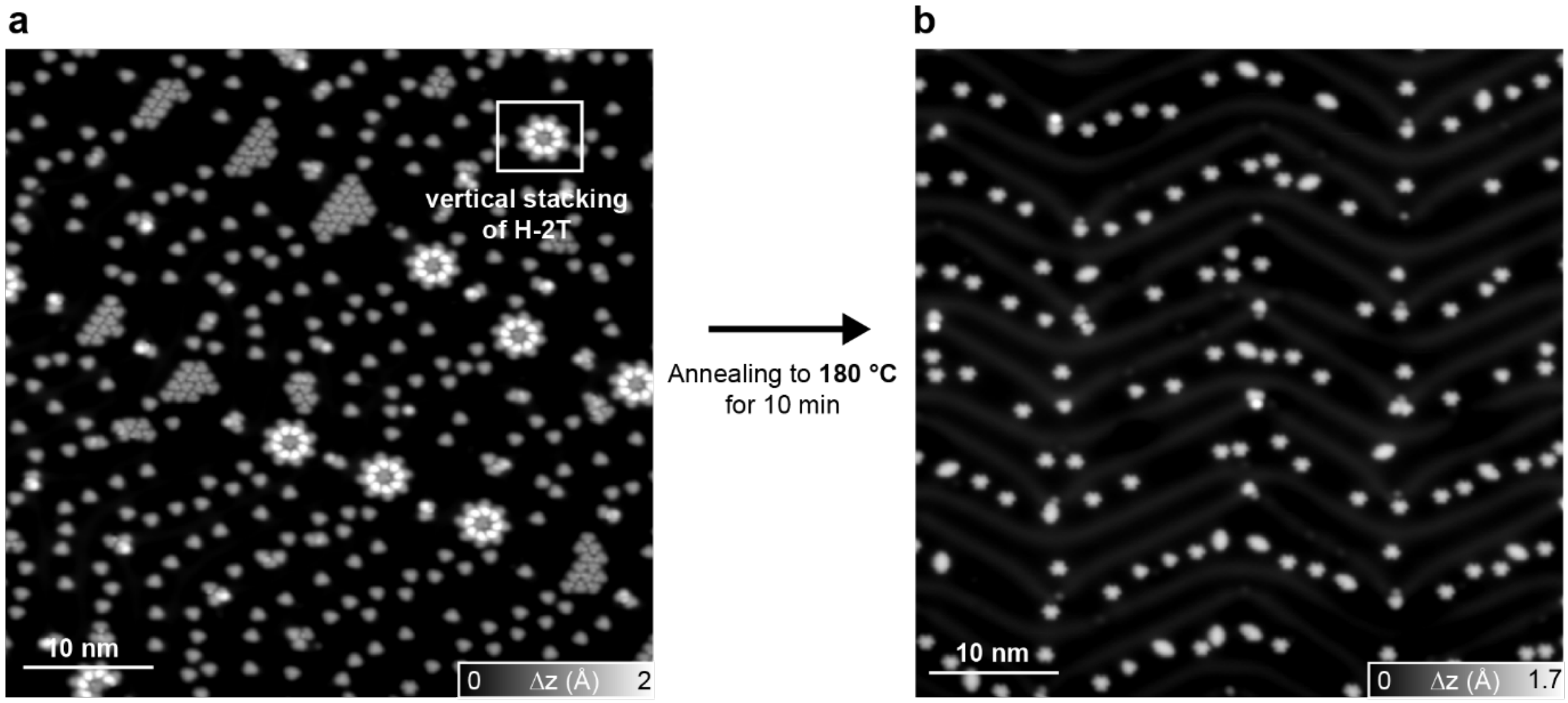

**Figure S2.** (**a**) Overview STM images of high-coverage sample of H-2T. (**b**) Thermally induced dehydrogenation of H-2T into the target 2T by annealing the Au(111) surface at 180 °C for 10 min. Notably, the annealing step reduces the surface coverage substantially due to monomer desorption. From the remaining molecules, ca. 70% of H-2T are converted to 2T. A minority of 2T has fused into structures with apparent rhombic shape, which we assign to peropyrenes. Scanning parameters (a,b): $V = -0.6$ V, $I = 100$ pA.

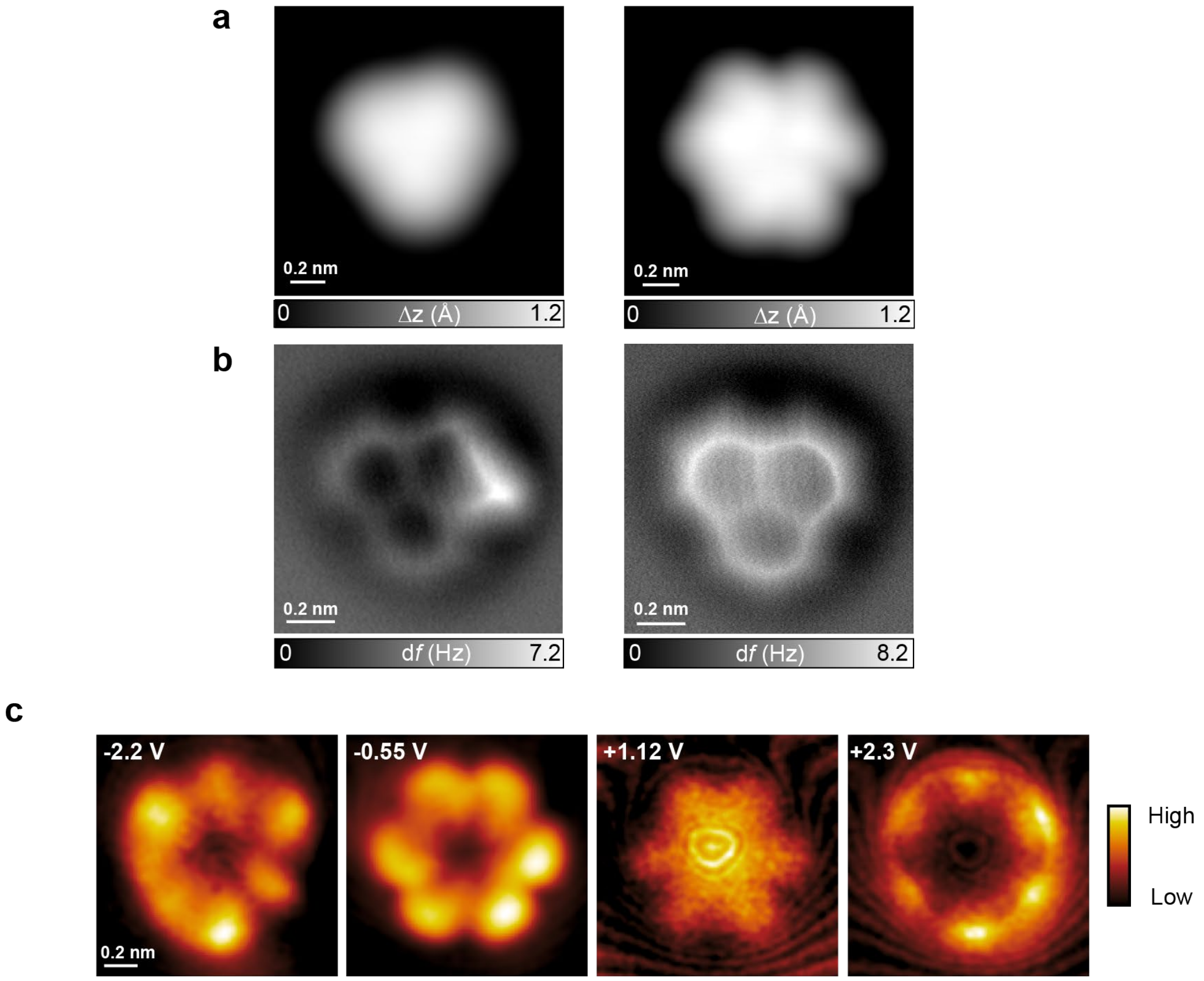

**Figure S3.** (**a**) High-resolution STM images for H-2T (left) and 2T (right) acquired with a CO-functionalized tip ($V = -100$ mV, $I = 100$ pA). (**b**) Corresponding nc-AFM image of H-2T and 2T (from left to right). Open feedback parameters: $V = -50$ mV, $I = 100$ pA; $\Delta z = -1.8$ Å (H-2T) and $-1.5$ Å (2T). (**c**) Constant-current d$I$/d$V$ maps of the molecular orbital resonances, acquired with a CO-functionalized tip. All the d$I$/d$V$ maps were acquired with a lock-in modulation $V_{rms} = 16$ mV and a current setpoint $I = 300$ pA.

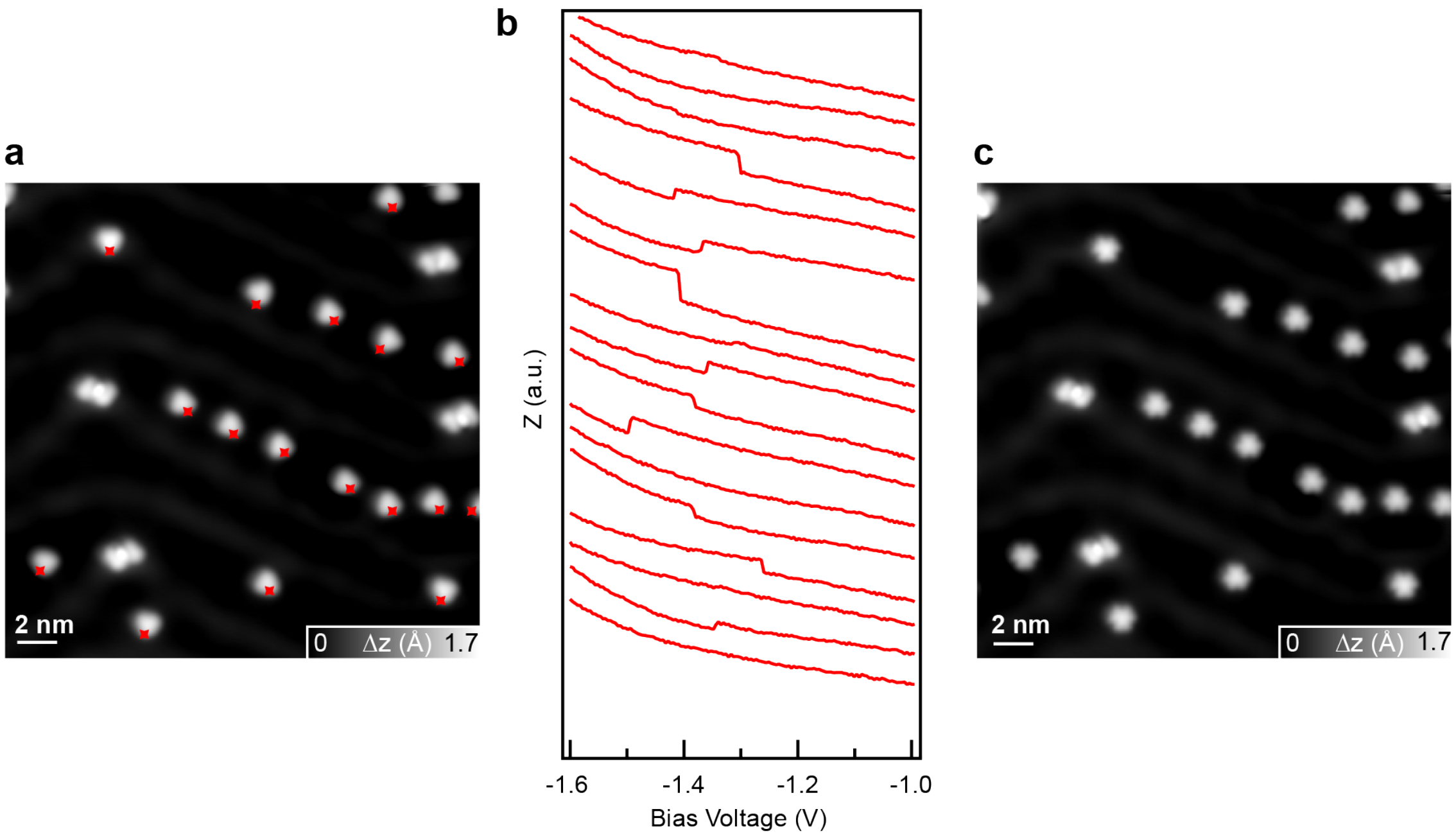

**Figure S4.** (**a**,**b**) Statistics on the tip-induced activation of H-2T molecules by $z(V)$ spectroscopy. (**a**) STM image of H-2T on Au(111) ($V = -100$ mV, $I = 100$ pA) where the $z(V)$ spectroscopy positions are depicted by red markers. (**b**) Cleaving of the hydrogen is detected as a step in $z(V)$ spectroscopy ($I = 50$ pA). Some spectra do not show a clear step. However, STM imaging (c) after the series of $z(V)$ spectra demonstrates the successful dehydrogenation of all targeted H-2T molecules into 2T ($V = -100$ mV, $I = 100$ pA).

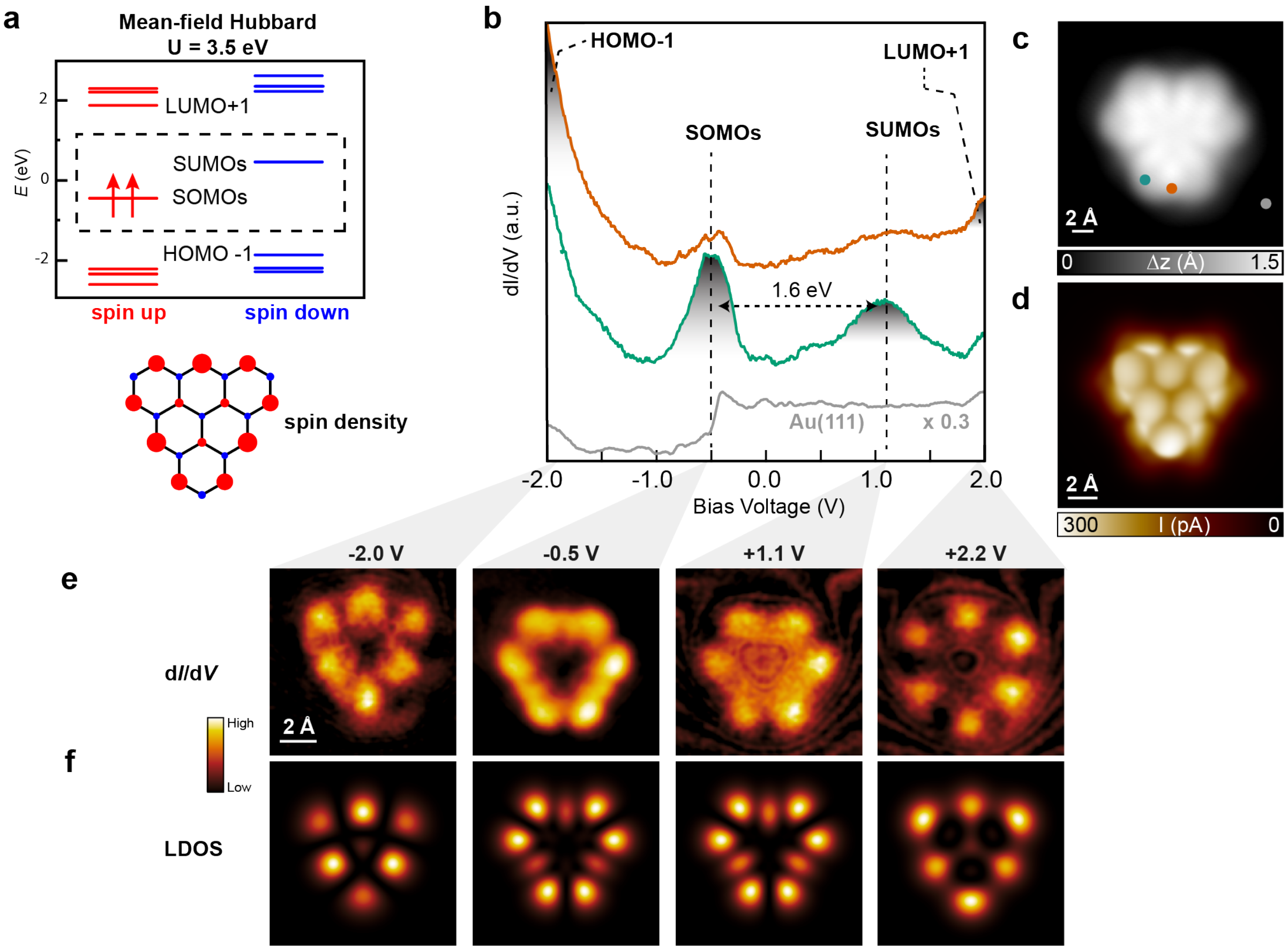

**Figure S5.** Electronic characterization of 3T. (**a**) MFH energy spectrum of 3T and spin polarization plot, where blue and red filled circles denote mean populations of spin up and spin down electrons, respectively. (**b**) d$I$/d$V$ spectroscopy on 3T revealing molecular orbital resonances (open feedback parameters: $V = -2.0$ V, $I = 300$ pA; $V_{rms} = 16$ mV). Acquisition positions are indicated in the HR-STM image shown in (c) ($V = -2.0$ V, $I = 100$ pA). (**d**) Bond-resolved STM image ($V = -5$ mV, $I = 50$ pA, $\Delta h = -0.75$ Å) of 3T on Au(111), acquired with a carbon monoxide (CO) functionalized tip. (**e**) Constant-current d$I$/d$V$ maps of the molecular orbital resonances, acquired with a CO-functionalized tip. All the d$I$/d$V$ maps were acquired with a lock-in modulation $V_{rms}$ = 16 mV and a current setpoint $I$ = 200 pA. (**f**) MFH-TB LDOS of HOMO-1, SOMOs, SUMOs, LUMO+1.

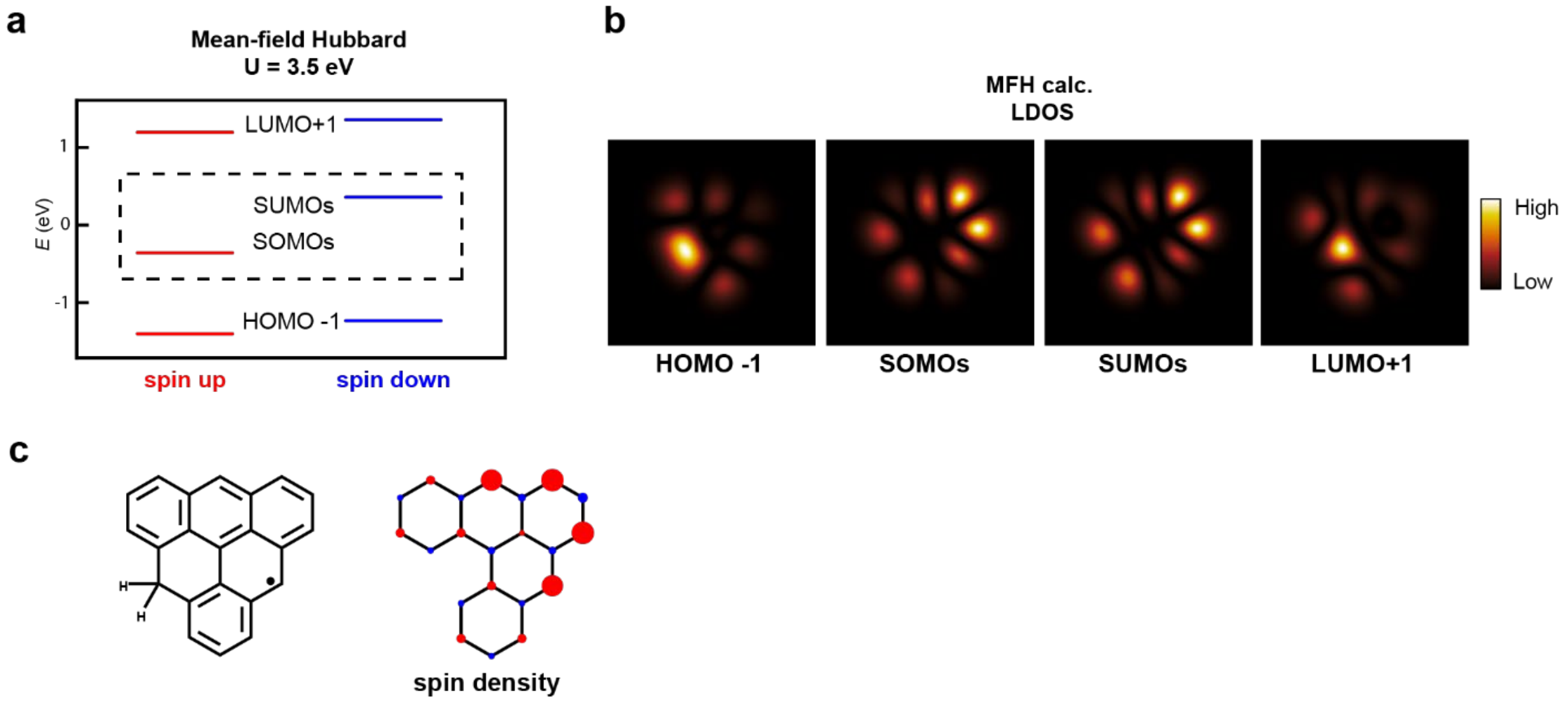

**Figure S6** MFH-TB calculations of H-3T. (**a**) MFH energy spectrum of H-3T and spin polarization plot, where blue and red filled circles denote mean populations of spin up and spin down electrons, respectively. The presence of the $sp^3$-hybridized carbon is taken in account by removing the respective carbon atom in the MFH-TB calculation. (**b**) MFH-TB LDOS of HOMO-1, SOMO, SUMO and LUMO+1 resonances of H-3T. The frontier states LDOS matches well with the apparent shape of the HR-STM image (Fig. 4b).

| Frota fit | 2T | H-3T | 3T |
|---|---|---|---|
| Γ (meV) | 6.1 ± 0.1 | 4.7 ± 0.1 | 4.3 ± 0.3 |
| $x_0$ (meV) | (3.0 ± 0.2)·10$^{-5}$ | (4.6 ± 0.3)·10$^{-4}$ | (0.27 ± 1.1)·10$^{-5}$ |
| *a* | 0.87 ± 0.01 | 1.11 ± 0.05 | 0.23 ± 0.01 |
| *b* | -0.90 ± 0.1 | -0.66 ± 0.01 | -0.53 ± 0.05 |
| *c* | -0.25 ± 0.01 | 0.31 ± 0.01 | -0.069 ± 0.001 |
| ϕ | 0.95 ± 0.01 | 0.97 ± 0.01 | 0.92 ± 0.01 |

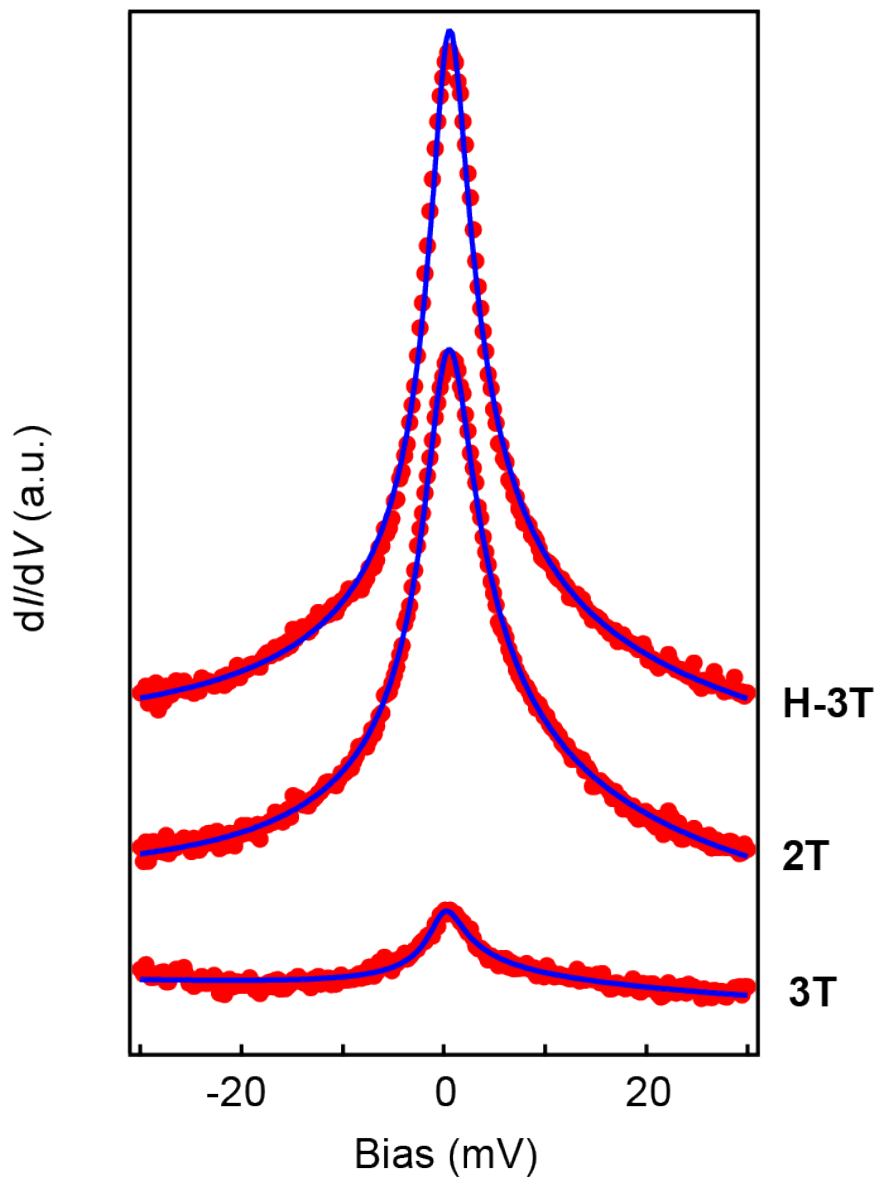

**Figure S7** Frota fitting parameters. The solid blue lines show the best fit of 2T, H-3T and 3T ZBRs with the Frota function[2] $-a \cdot Re\left[e^{i\pi\phi}\sqrt{\frac{i\cdot 0.39\Gamma}{eV-x_0+i\cdot 0.39\Gamma}}\right]+b\cdot V+c$. Where $a$, $b$, $c$ parameters represent the amplitude factor, a linear slope of the d*I*/d*V* curve and a conductance offset, respectively. $\phi$ is a phase factor defining the shape of the Kondo resonance, $x_0$ and Γ are respectively the energy position of the resonance and its half width at half maximum (HWHM). The fitting parameters of each system are listed in the table.

## 3. Synthetic procedures

### Solution synthesis of H-2T

1*H*-phenalene (H-2T) was prepared from commercially available 3-(naphthalen-1-yl)propanoic acid (1) in three steps. The phenalene-core was built up by a Friedel–Crafts acylation. The reduction of 2,3-dihydro-*1H*-phenalen-1-one (2) and a subsequent dehydration of 2,3-dihydro-1*H*-phenalen-1-ol (3) yielded H-2T. The air sensitive compound was purified by column chromatography and subsequent sublimation. Hereafter a detailed description of the synthesis steps of H-2T is reported.

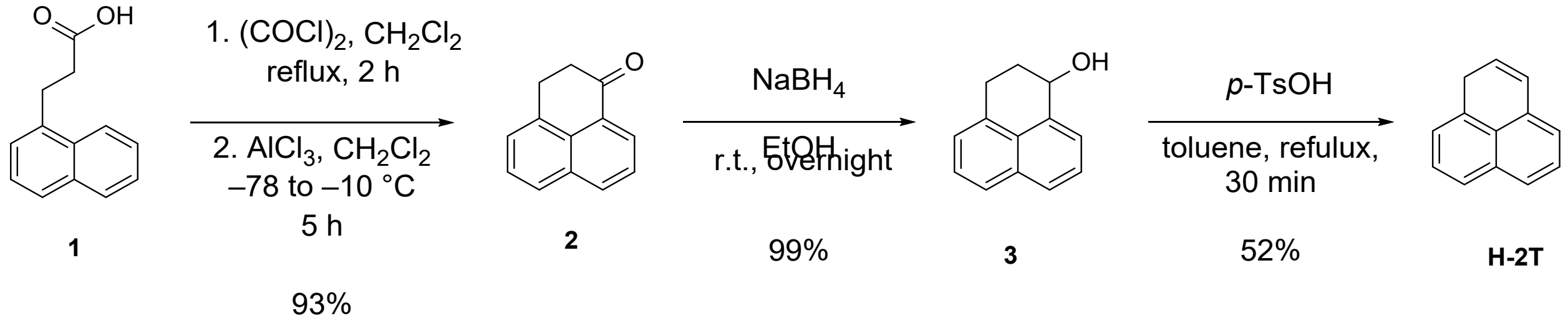

**2,3-Dihydro-1*H*-phenalen-1-one (2).**[5] The reaction was carried out under inert conditions. A solution of 3-(naphthalen-1-yl)propanoic acid (**1**, 4.50 g, 22.5 mmol) in oxalyl chloride (60 mL) was refluxed for 2 h. After excessive oxalyl chloride was removed under reduced pressure, the crude product was dissolved in $CH_2Cl_2$ (60 mL) and cooled to –78 °C. $AlCl_3$ (4.5 g, 33.8 mmol) was added to the solution before the reaction mixture was allowed to warm up to –10 °C over a period of 5 h. The reaction mixture was subsequently poured on ice and the organic layer was separated before the aqueous layer was extracted with $CH_2Cl_2$. The combined organic layers were washed with aq. $NaHCO_3$ (sat.) and were dried over $MgSO_4$. After evaporation of the solvents, the crude product was purified by flash column chromatography ($SiO_2$, $CH_2Cl_2$) to afford the desired compound as an off-white solid (3.80 g, 20.9 mmol, 93%). [CAS 518-85-4] **$^1$H NMR (400 MHz, $CDCl_3$, ppm):** $\delta$ 8.21 (dd, $J$ = 7.2, 1.3 Hz, 1H), 8.10 (dd, $J$ = 8.2, 1.3 Hz, 1H), 7.81 (dd, $J$ = 7.9, 1.6 Hz, 1H), 7.61 (dd, $J$ = 8.2, 7.2 Hz, 1H), 7.55–7.43 (m, 2H), 3.49–3.40 (m, 2H), 2.99 (dd, $J$ = 7.9, 6.3 Hz, 2H). **$^{13}$C NMR (101 MHz, $CDCl_3$, ppm):** $\delta$ 198.7, 134.2, 133.5, 133.4, 131.8, 130.0, 126.4, 125.8, 125.7, 125.2, 38.7, 28.7. **HRMS (EI) *m/z*:** $[M]^+$ Calcd for $C_{13}H_{10}O$ 182.0726; Found 182.0724.

**2,3-Dihydro-1*H*-phenalen-1-ol (3).**[5] To a solution of 2,3-dihydro-1*H*-phenalen-1-one (**2**, 300 mg, 1.65 mmol) in EtOH (7 mL) was added $NaBH_4$ (98.7 mg, 1.82 mmol) in one portion. The reaction mixture was stirred at rt overnight before it was quenched with water. The reaction mixture was extracted with $CH_2Cl_2$, the organic layers were washed with aq. $NaHCO_3$ (sat.), dried over $MgSO_4$ and the solvent was evaporated. The crude product was purified by flash column chromatography ($SiO_2$, cyclohexane/EtOAc 4:1 v/v) to afford the desired compound as a colorless solid (304 mg, 1.65 mmol, 99%). [CAS 130292-28-3] The obtained $^1$H-NMR spectrum is in accord with the previously reported data[6]. **$^1$H NMR (400 MHz, $CDCl_3$, ppm):** *δ* 7.80 (dd, *J* = 8.2, 1.3 Hz, 1H), 7.72 (dd, J = 8.2, 1.1 Hz, 1H), 7.57 (dt, *J* = 7.0, 1.2 Hz, 1H), 7.48 (dd, *J* = 8.2, 7.0 Hz, 1H), 7.42 (dd, *J* = 8.2, 6.9 Hz, 1H), 7.31 (dq, *J* = 7.0, 1.3 Hz, 1H), 5.12 (dd, *J* = 6.6, 3.8 Hz, 1H), 3.40–3.04 (m, 2H), 2.29–2.13 (m, 2H), 1.81 (s, 1H). **$^{13}$C NMR (75 MHz, $CDCl_3$, ppm):** *δ* 137.5, 135.2, 133.8, 128.8, 128.2, 126.0, 125.8, 125.7, 124.5, 123.6, 69.5, 31.4, 26.3. **HRMS (EI) *m/z*:** $[M]^+$ Calcd for $C_{13}H_{12}O$ 184.0883, Found 184.0883.

**1*H*-Phenalene (H-2T).**[5] The reaction and workup were carried out under inert conditions, solvents were deoxygenated. A solution of 2,3-dihydro-1*H*-phenalen-1-ol (**3,** 151 mg, 0.82 mmol) in toluene (5 mL) was refluxed, before a catalytic amount of *p*-TsOH was added. The reaction mixture was refluxed for further 0.5 h. Then, the reaction mixture was cooled to room temperature und concentrated in vacuum. The residing liquid was diluted with petroleum ether and passed through a pad of Florisil with petroleum ether as eluent. The solvent was removed in vacuum and the residing solid was sublimed (~110 °C, $2*10^{-2}$ mbar) to afford the desired compound as a colorless solid (70.3 mg, 0.42 mmol, 52%). CAS [203-80-5] **MP.:** 84–96 °C (Lit. 83–84 °C)[7]. The characterization of 1*H*-phenalene in literature is incomplete, we here report $^1$H and $^{13}$C resonances in $CD_2Cl_2$ and its high-resolution mass spectrum. **$^1$H NMR (400 MHz, $CD_2Cl_2$, ppm):** *δ* 7.55 (dd, *J* = 8.1, 1.4 Hz, 1H), 7.52 (dd, *J* = 8.4, 1.1 Hz, 1H), 7.39–7.32 (m, 1H), 7.27 (dd, *J* = 8.3, 6.9 Hz, 2H), 6.98 (dd, *J* = 6.9, 1.3 Hz, 1H), 6.60 (dt, *J* = 9.9, 2.2 Hz, 1H), 6.06 (dt, *J* = 9.9, 4.1 Hz, 1H), 4.10–4.03 (m, 2H). **$^{13}$C NMR (101 MHz, $CD_2Cl_2$, ppm):** *δ* 134.7, 134.0, 132.4, 129.8, 128.4, 127.8, 127.0, 126.6, 126.5, 125.3, 122.4, 32.4. **HRMS (EI) m/z:** $[M]^+$ Calcd for $C_{13}H_{10}$ 166.0778, Found 166.0766.

# 4. High-resolution mass spectra and NMR characterization

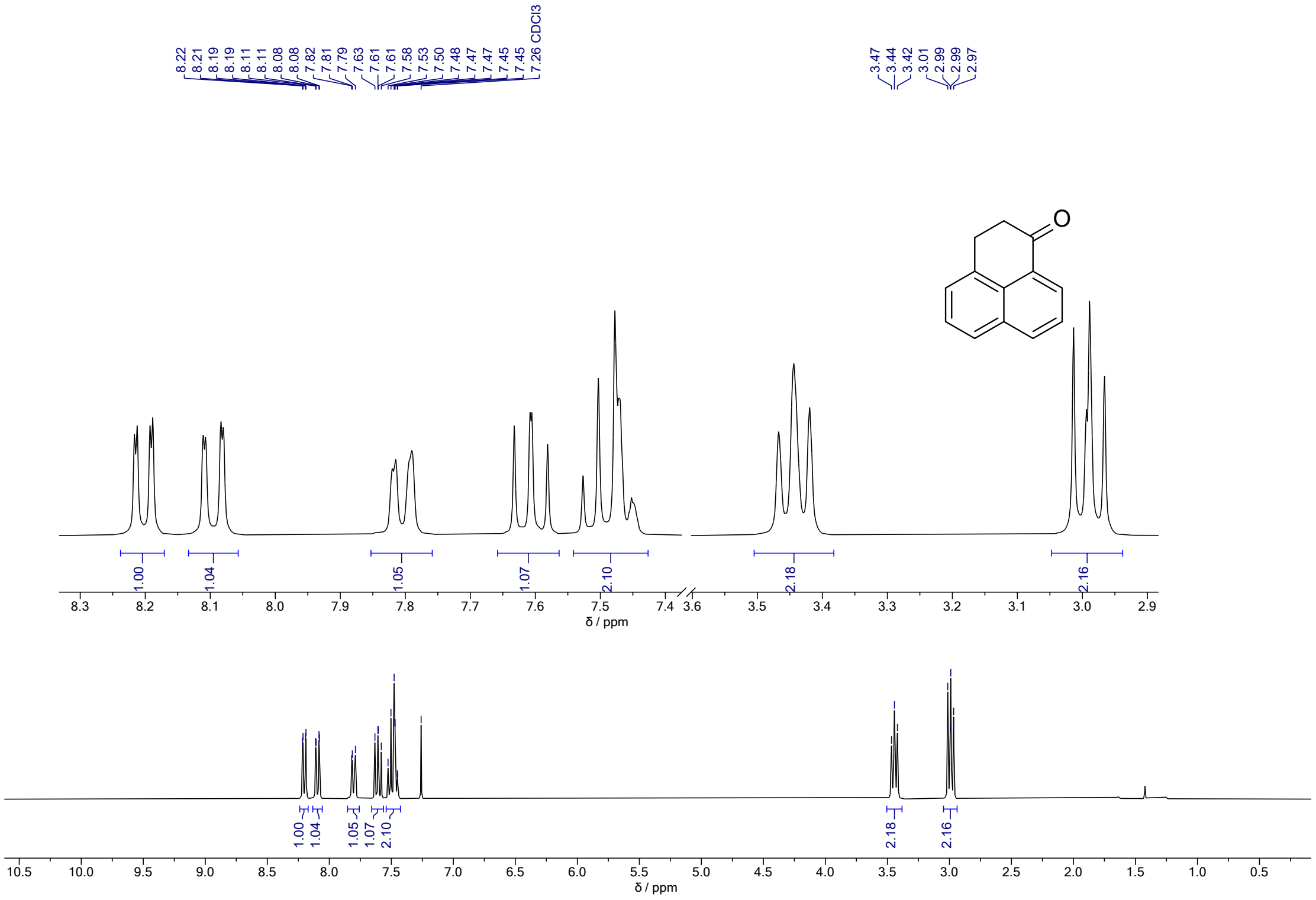

$^{13}$C NMR / 101 MHz / $CDCl_3$ / 25 °C

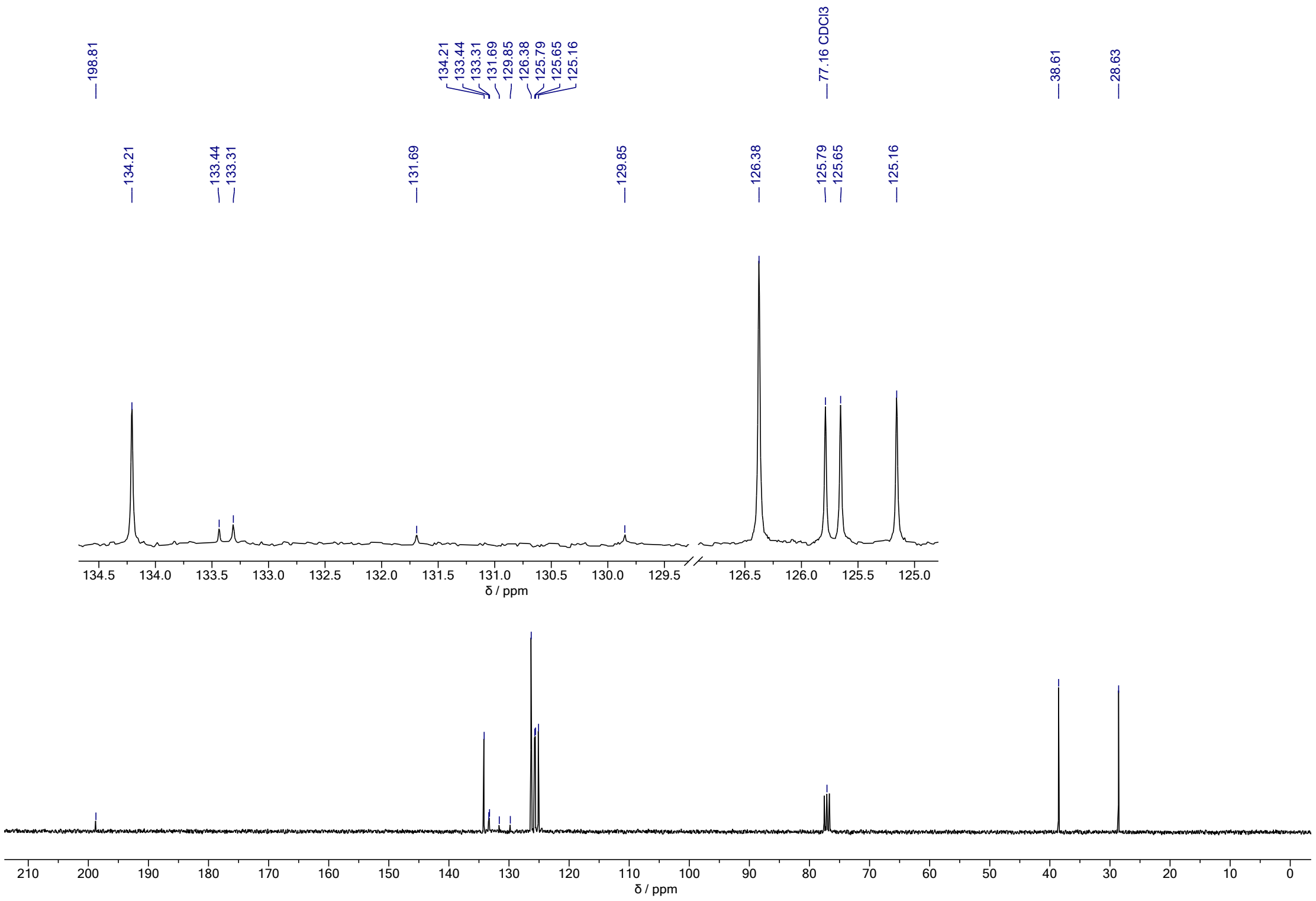

2,3-Dihydro-1*H*-phenalen-1-ol (151) $^{1}$H NMR / 400 MHz / $CDCl_3$ / 25 °C

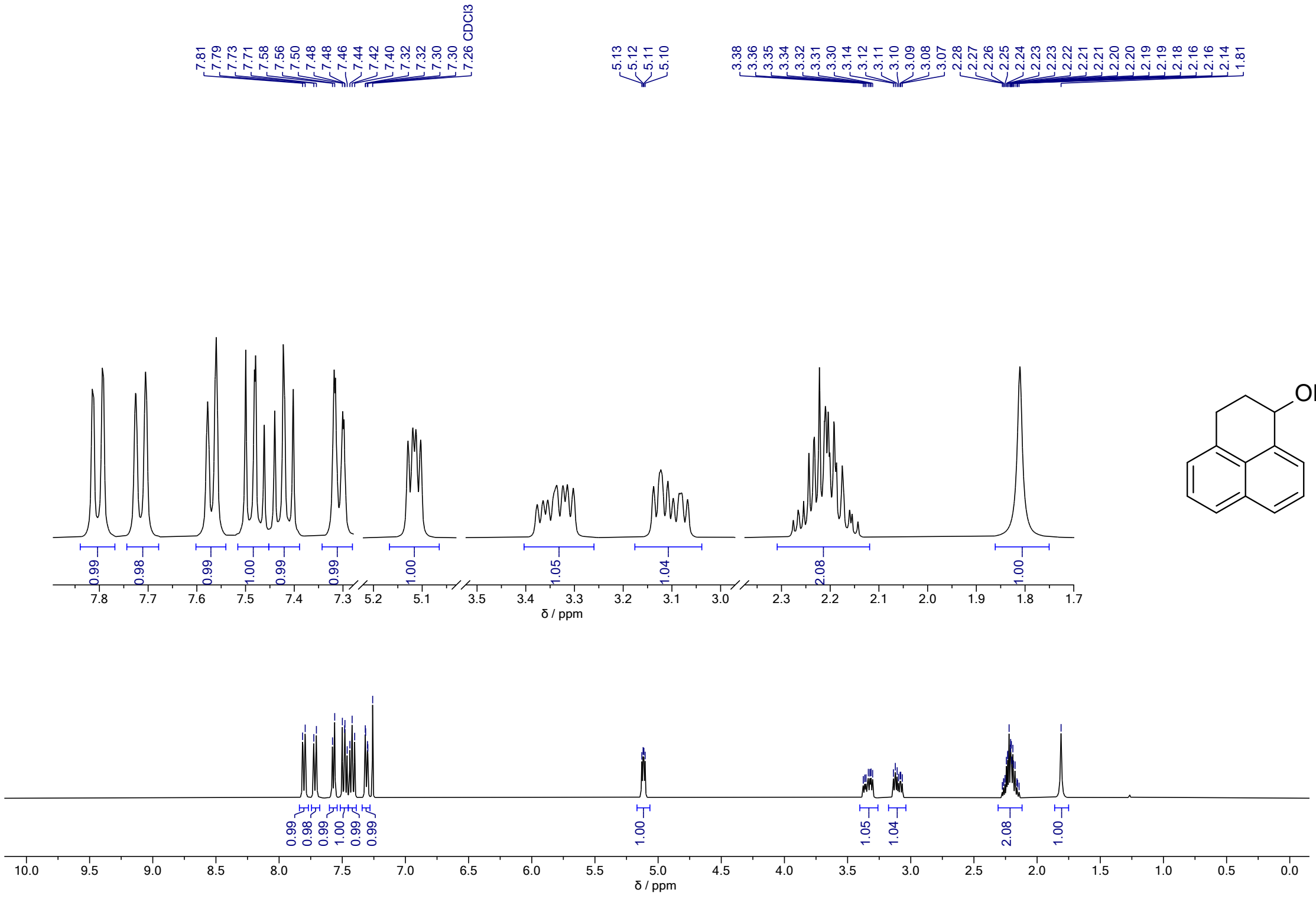

$^{13}C$ NMR / 101 MHz / $CDCl_3$ / 25 °C

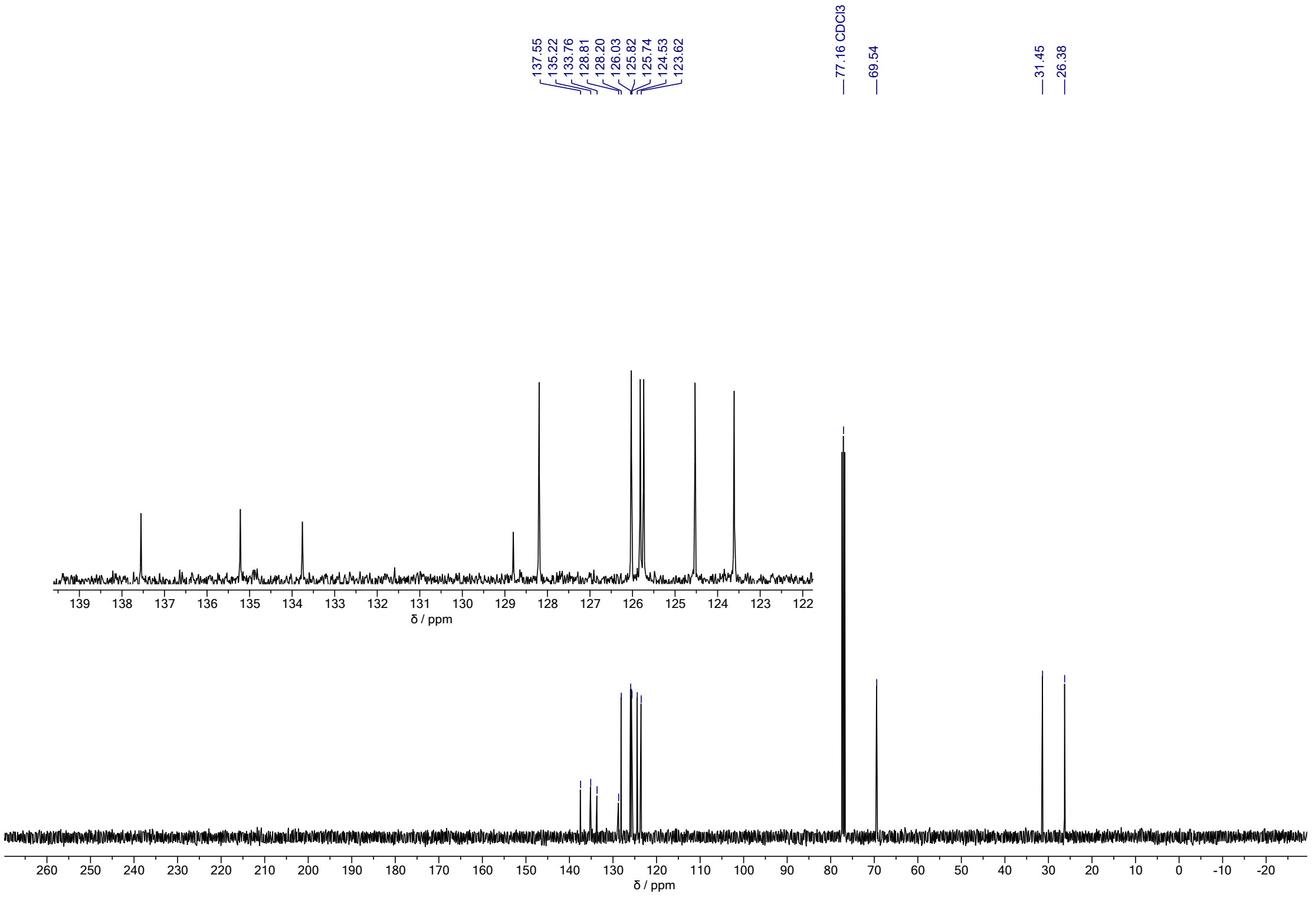

## HRMS (EI)

HR-EI-Report (Thermo DFS)

Sample: Client:

7151juhr-cmass2 #87-92 RT: 12.10-12.14 AV: 6 SB: 8 12.03-12.09NL:2.11E6 T:
+ c EI Full ms [178.29-195.29]

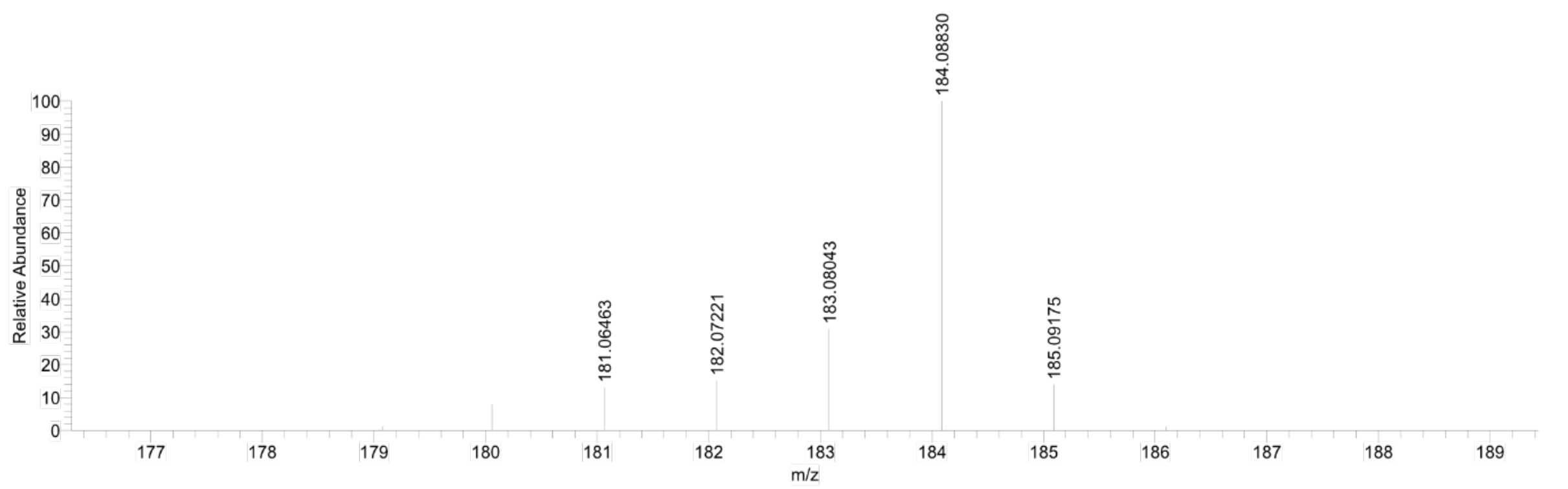

7151juhr-cmass2#87-92 RT: 12.10-12.14 AV: 6 T: + c
EI Full ms [178.29-195.29]
m/z= 180.78581-184.29280

| m/z | Intensity | Relative | Theo. Mass | Delta (ppm) | Composition |
|---|---|---|---|---|---|
| 181.06463 | 275894.8 | 13.05 | 181.06479 | -0.88 | C 13 H 9 O |
| 182.07221 | 320293.7 | 15.15 | 182.07262 | -2.23 | C13 H10 O |
| 183.08043 | 652154.2 | 30.86 | 183.08044 | -0.07 | C13 H11 O |
| 184.08830 | 2113606.0 | 100.00 | 184.08827 | 0.18 | C13 H12 O |

**1*H*-Phenalene (152)** $^{1}$H NMR / 400 MHz / $CD_2Cl_2$ / 25 °C

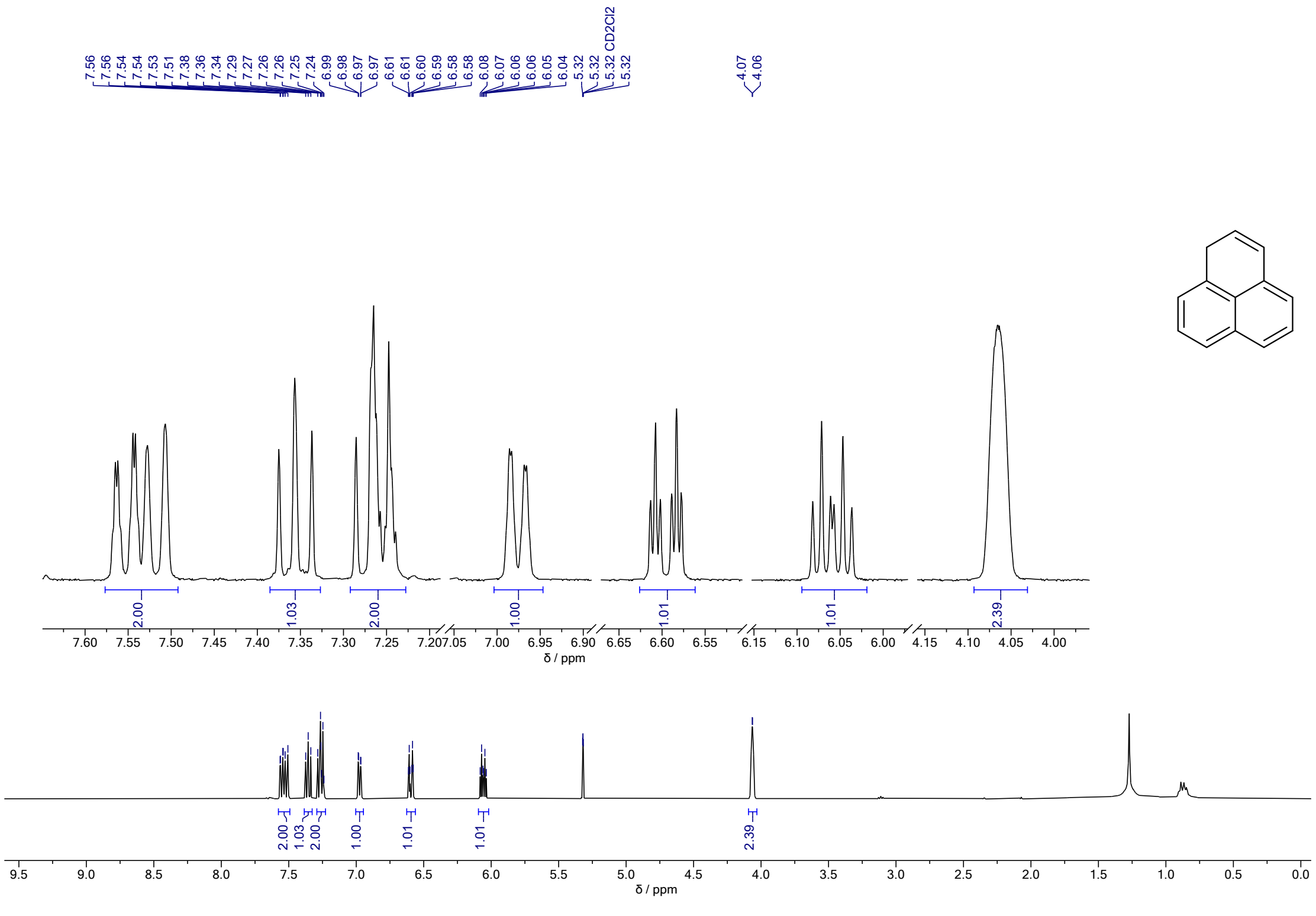

$^{13}C$ NMR / 101 MHz / $CD_2Cl_2$ / 25 °C

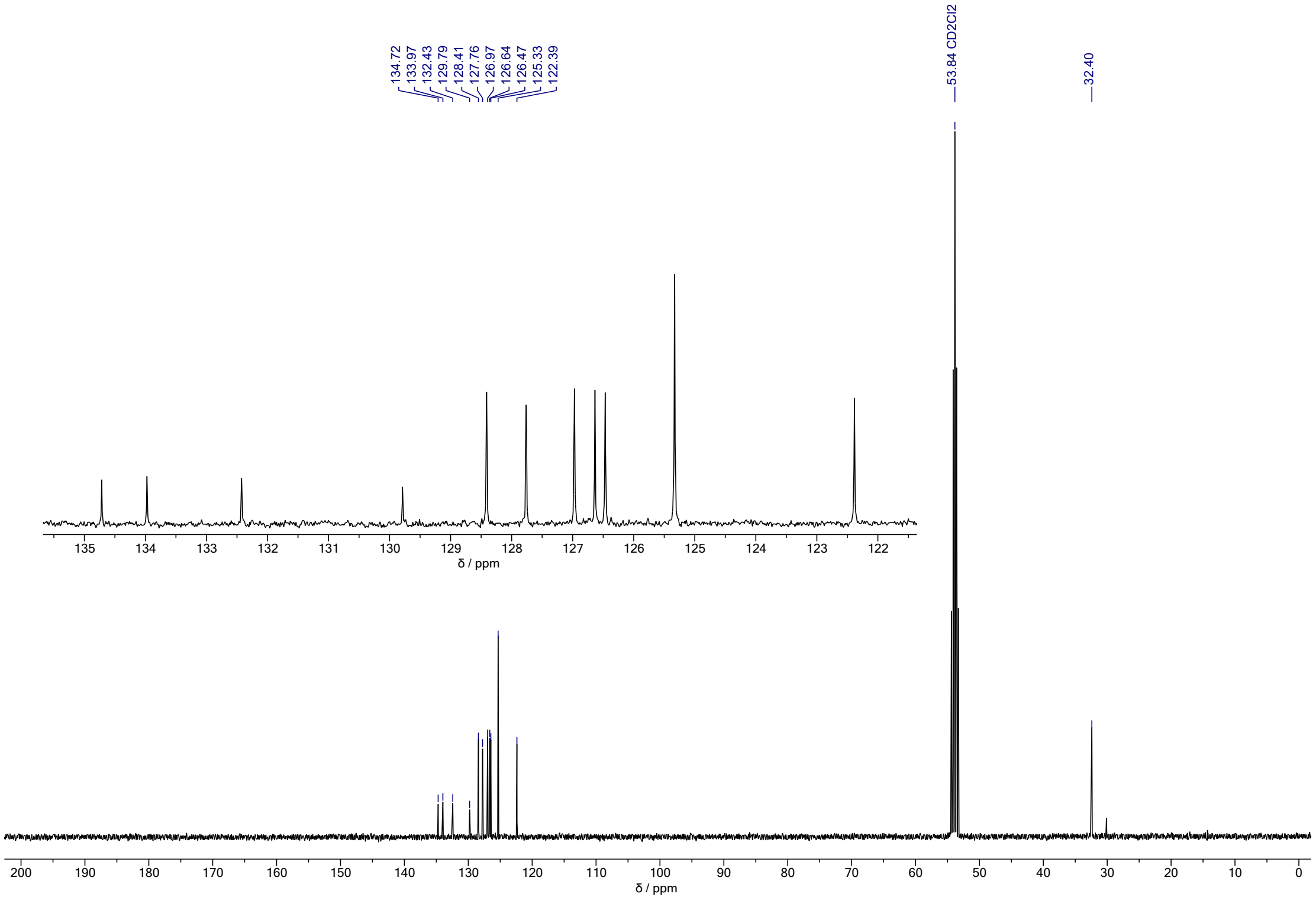

HRMS (EI)

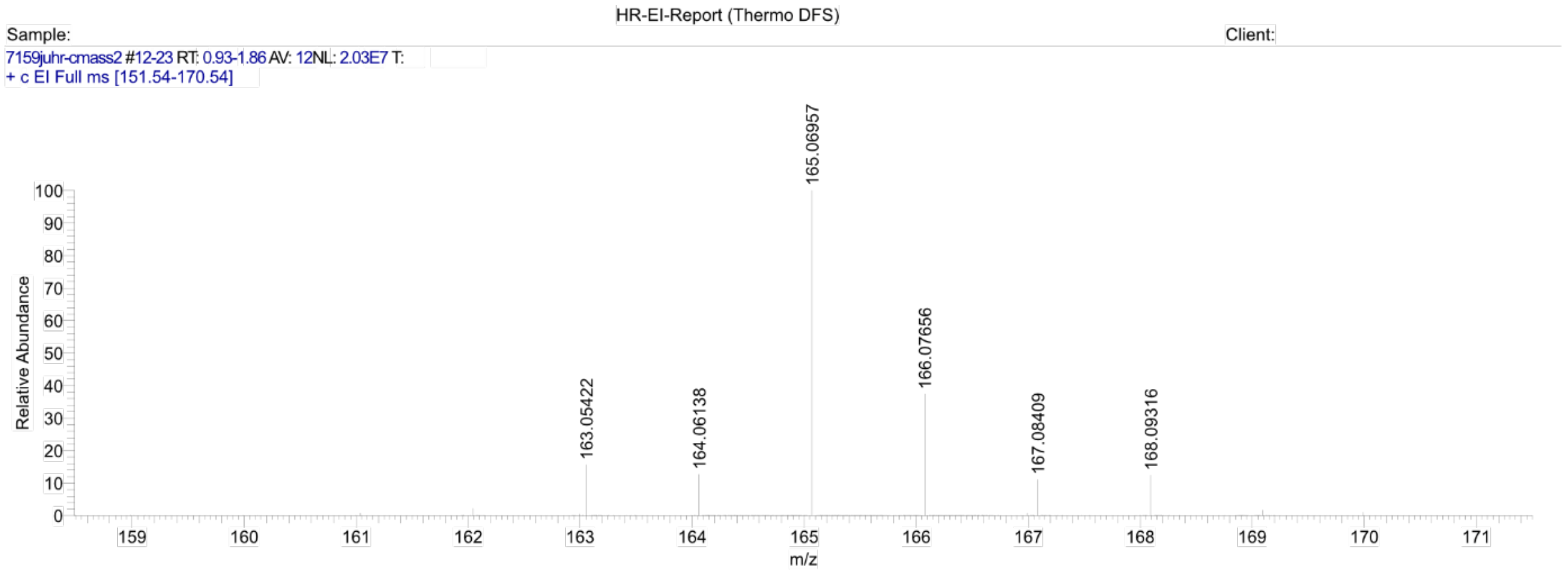

7159juhr-cmass2#12-23 RT: 0.93-1.86 AV: 12 T: +
c EI Full ms [151.54-170.54]
m/z= 162.40806-168.34751

| m/z | Intensity | Relative | Theo. Mass | Delta (ppm) | Composition |
|---|---|---|---|---|---|
| 163.05422 | 3167051.6 | 15.64 | 163.05423 | -0.05 | $C_{13}H_7$ |
| 164.06138 | 2569084.3 | 12.69 | 164.06205 | -4.07 | $C_{13}H_8$ |
| 165.06957 | 20250922.7 | 100.00 | 165.06988 | -1.83 | $C_{13}H_9$ |
| 166.07656 | 7566182.8 | 37.36 | 166.07770 | -6.88 | $C_{13}H_{10}$ |
| 167.08409 | 2248272.6 | 11.10 | 167.08553 | -8.60 | $C_{13}H_{11}$ |
| 168.09316 | 2509202.2 | 12.39 | 168.09335 | -1.17 | $C_{13}H_{12}$ |